\documentclass[pre,twocolumn,amssymb]{revtex4}
\usepackage{amsmath}
\usepackage{graphicx}
\usepackage{color}
\usepackage{epsfig}
\usepackage{latexsym}
\usepackage{bm}
\usepackage{ulem}
\usepackage{dutchcal}
\usepackage{subfigure}

\begin{document}
\renewcommand{\ni}{{\noindent}}
\newcommand{\dprime}{{\prime\prime}}
\newcommand{\be}{\begin{equation}}
\newcommand{\ee}{\end{equation}}
\newcommand{\bea}{\begin{eqnarray}}
\newcommand{\eea}{\end{eqnarray}}
\newcommand{\nn}{\nonumber}
\newcommand{\bk}{{\bf k}}
\newcommand{\bQ}{{\bf Q}}
\newcommand{\q}{{\bf q}}
\newcommand{\s}{{\bf s}}
\newcommand{\bN}{{\bf \nabla}}
\newcommand{\bA}{{\bf A}}
\newcommand{\bE}{{\bf E}}
\newcommand{\bj}{{\bf j}}
\newcommand{\bJ}{{\bf J}}
\newcommand{\bs}{{\bf v}_s}
\newcommand{\bn}{{\bf v}_n}
\newcommand{\bv}{{\bf v}}
\newcommand{\la}{\langle}
\newcommand{\ra}{\rangle}
\newcommand{\dg}{\dagger}
\newcommand{\br}{{\bf{r}}}
\newcommand{\brp}{{\bf{r}^\prime}}
\newcommand{\bq}{{\bf{q}}}
\newcommand{\hx}{\hat{\bf x}}
\newcommand{\hy}{\hat{\bf y}}
\newcommand{\bS}{{\bf S}}
\newcommand{\cU}{{\cal U}}
\newcommand{\cD}{{\cal D}}
\newcommand{\bR}{{\bf R}}
\newcommand{\pll}{\parallel}
\newcommand{\sumr}{\sum_{\vr}}
\newcommand{\cP}{{\cal P}}
\newcommand{\cQ}{{\cal Q}}
\newcommand{\cS}{{\cal S}}
\newcommand{\ua}{\uparrow}
\newcommand{\da}{\downarrow}
\newcommand{\red}{\textcolor {red}}
\newcommand{\blu}{\textcolor {blue}}

\def\lsim {\protect \raisebox{-0.75ex}[-1.5ex]{$\;\stackrel{<}{\sim}\;$}}
\def\gsim {\protect \raisebox{-0.75ex}[-1.5ex]{$\;\stackrel{>}{\sim}\;$}}
\def\lsimeq {\protect \raisebox{-0.75ex}[-1.5ex]{$\;\stackrel{<}{\simeq}\;$}}
\def\gsimeq {\protect \raisebox{-0.75ex}[-1.5ex]{$\;\stackrel{>}{\simeq}\;$}}


\title{Hydrodynamics, superfluidity and giant number fluctuations in a model of self-propelled particles  }

\author{Tanmoy Chakraborty$^1$, Subhadip Chakraborti$^{1,2}$, Arghya Das$^{2}$, and Punyabrata Pradhan$^1$}

\affiliation{$^1$Department of Theoretical Sciences, S. N. Bose National Centre for Basic Sciences, Block-JD, Sector-III, Salt Lake, Kolkata 700106, India \\ $^2$International Centre for Theoretical Sciences, Tata Institute of Fundamental Research, Bengaluru 560089, India }

\begin{abstract}

\noindent{We derive hydrodynamics of a prototypical one dimensional model, having variable-range hopping, which mimics passive diffusion and ballistic motion of active, or self-propelled, particles. The model has two main ingredients - the hardcore interaction and the competing mechanisms of short and long range hopping.
We calculate  two density-dependent transport coefficients - the bulk-diffusion coefficient and the conductivity, the ratio of which, despite violation of detailed balance, is connected to number fluctuation by an Einstein relation. 
In the limit of infinite range hopping, the model exhibits, upon tuning density $\rho$ (or activity), a ``superfluid'' transition from a finitely conducting state to an infinitely conducting one, characterized by a divergence in conductivity $\chi(\rho) \sim (\rho-\rho_c)^{-1}$ with $\rho_c$ being the critical density. The diverging conductivity greatly increases particle (or vacancy) mobility and induces ``giant'' number fluctuations in the system.    
}
\typeout{polish abstract}
\end{abstract}

\maketitle

\section{Introduction} 

Persistence and interactions are the hallmarks of self-propelled particles (SPPs), also called active matters. Self-propelled particles are ubiquitous in nature - in living beings, e.g., bacterial colony \cite{Goldstein_PRL2004, Ariel_NatCom2015, Cox, Matthaus, Tu_Grinstein}, flocking birds \cite{birds} and fish schools \cite{fish} as well as in nonliving systems, e.g., photo-activated colloids \cite{Palacci}; for review, see \cite{Sriram_RMP2013}. 
They propel themselves persistently by consuming chemical energy, while interacting with their neighbors through chemical signalling or excluded-volume interactions, and dissipate energy to the medium. Due to the subtle interplay between drive, dissipation and interactions, SPPs remain inherently out of equilibrium and exhibit fascinating collective behaviors like clustering \cite{Bar_PRE2006, Peruani_PRL2012, Marchetti_PRL2012, Baskaran_PRL2013, Levis_PRE2014, Barberis, Evans_PRL2018} and ``giant'' number fluctuations (GNF) \cite{Peruani_PRL2012, Marchetti_PRL2012, Sriram_EPL2003, Menon_Science2007, Swinney_PNAS2010, Rajesh_PRL2012, Chate_PRE2017} on the one hand and anomalous transport on the other \cite{Hatwalne_PRL2004, Cates_PRL2008, Yeomans_PRL2008, Prost_EPL2005, Sokolov_PRE2007, Sokolov_PRL2009, Dunkel_PRL2013, Benichou_PRL2013, Clement, Golestanian_PRE2019}.

There has been considerable progress in understanding collective behaviors of SPPs through studies of simple models, such as Vicsek model \cite{Vicsek_PRL1995, Toner, Ihle}, run-and-tumble particles (RTPs) \cite{Cates_PRL2008, Soto_PRE2014, Levis_PRE2014, Blythe_PRL2016}, active Brownian particles \cite{Marchetti_PRL2012, Baskaran_PRL2013}, and active lattice gases \cite{Tailleur_PRL2013, Whitelam_JCP2017, Tailleur_PRL2018}. However, even for these minimal models, exact results are few and far between \cite{Toner, Tailleur_PRL2018}, mainly because such systems are not in equilibrium, have nontrivial many-body correlations and the probability weights of their microscopic configurations are unknown. Not surprisingly, there is lack of concrete understanding of two important questions: (i) What are precisely the underlying mechanisms responsible for the anomalous behaviors in SPPs and (ii) how are fluctuations and transport related? In this paper, we address these issues in a minimalistic setting of a prototypical many-particle model, which qualitatively captures the large-scale features of SPPs and, moreover, is amenable to exact analysis.

Indeed, an exact derivation of hydrodynamics of interacting SPPs, accounting for long-ranged spatio-temporal correlations as manifest in the anomalous behaviors of fluctuation and transport, has been elusive so far. To bridge this gap, we introduce a generalized version of simple symmetric exclusion process (SSEP) \cite{SSEP}, called generalized long-hop  model (gLHM), which, in addition to the nearest-neighbor short-range hopping of SSEP, incorporates also long-range hopping by particles. The model is motivated by random, but space-time correlated coherent motion, called ``runs'' or ``swims'', which are observed in living micro-organisms such as bacteria and amoebae \cite{Cox, Ariel_NatCom2015, Matthaus}. For example, consider a bacterium like {\it E. Coli}, which moves by rotating its flagella: Coherent counter-clockwise rotation drives the {\it E. Coli} in a straight line by some distance and disassembled clockwise rotation makes bacterium tumble in a random direction \cite{Tu_Grinstein}. On a large time scale, the motion of {\it E. Coli} can be traced as a zigzag path consisting of series of ballistic ``swims'', punctuated by ``tumbles''.  However, depending on the fluctuations of an enzyme in a bacterium's chemotaxis network (e.g., CheY-P in {\it E. Coli} \cite{Tu_Grinstein}), there can be some variations in the individual bacterium's swim-lengths \cite{Matthaus}, having a typical characteristic length scale, called persistence length. In certain conditions though, the swim-length distributions  have long-tails with diverging mean, sharing the characteristics of Levy-walks \cite{Ariel_NatCom2015}.

In the generalized long-hop model (gLHM) introduced in this paper, we consider hardcore particles moving on a one-dimensional lattice on a ring, with total number of particles being conserved. Provided that there is an empty lane (an empty stretch of vacant sites) in front of it, a particle can hop a variable distance, symmetrically in either direction. The mean hop-length in gLHM could be related to the persistence length of individual self-propelled micro-organisms like bacteria. Though we model persistence in the simplest possible way, the model, as explained later, brings to the fore a crucial element, which could be central to the understanding of clustering and transport in self-propelled-particles: The competition between long and short range hopping mechanisms induces cooperative behaviors in the system. Indeed, it is not difficult to see that, while a particle makes a long range hop, equivalently a ``hole'' or a vacancy cluster as a whole moves in unison. Subsequently, two such neighboring clusters could then coalesce to form even a larger one, effectively incorporating cooperativity into the dynamics (see the model in Fig. \ref{model} and Sec. II).  Specifically in the context of micro-organisms, while the long-hops correspond to the persistent or the ballistic motion, the short-hops mimic thermal diffusion in the surrounding solvent.
The relative strength of long range hopping is called here activity.

The main results of the paper are the following. We derive, from first principles, the hydrodynamic structure of generalized long-hop model in  the diffusive scaling limit. Moreover, in a special case of the model with an infinite range hopping, we explicitly obtain in one dimension the analytic expressions of two transport coefficients - the bulk diffusion coefficient $D(\rho)$ and the conductivity, or the inverse resistivity, $\chi(\rho)$. The transport coefficients in general are nonlinear functions of density $\rho$ and are defined through the diffusive current $J_D= - D(\rho) \partial \rho(x,t)/ \partial x$ and the drift current $J_d = \chi(\rho) F$, respectively, where $\rho(x,t)$ is the density at position $x$ and time $t$, and $F$ is the magnitude of a small external force field, which is applied to calculate the (linear) response of the system to the external perturbation. 
Remarkably, even in the absence of detailed balance, we find an Einstein relation $\sigma^2(\rho) = \chi(\rho)/D(\rho)$, which relates scaled subsystem particle-number fluctuation $\sigma^2(\rho) = \lim_{l{sub} \rightarrow \infty} (\langle n^2 \rangle - \langle n \rangle^2)/l_{sub}$ to the ratio of the two transport coefficients, where $n$ is the number of particles in a subsystem of size $l_{sub}$. 
Indeed, the competition between the short and the long range hopping induces, beyond a critical density $\rho_c$ (or a critical activity), a first-order ``superfluid'' transition from a finitely conducting state to an infinitely conducting one. In the ``superfluid'' phase, the particles (or vacancies) are highly mobile and consequently the conductivity diverges near criticality as $\chi(\rho) \sim (\rho - \rho_c)^{-1}$; in other words, the resistivity vanishes. This extremely high mobility near criticality leads to ``giant'' number fluctuations, which, along with the diverging conductivity, persist even in the ordered phase; interestingly, the bulk-diffusion coefficient remains finite.

The paper is organized as follows. In Sec. II, we define generalized long-hop model. In section III, we derive hydrodynamic structure of the model in terms of the bulk-diffusion coefficient and the conductivity: Hydrodynamics for finite range hopping in Sec. \ref{sec-frh} and for infinite range hopping in Sec. \ref{sec-IRH}; we verify density relaxation governed by the above hydrodynamics in Sec. \ref{sec-den-rel} and the existence of an Einstein relation in Sec. \ref{sec-Ein-Rel}, we discuss the connection between ``superfluid'' transition and ``giant'' number fluctuation in Sec. \ref{sec-gnf}. In Sec. \ref{sec-sum}, we summarize with some concluding remarks.

\section{Model}

\begin{figure}
\begin{center}
\leavevmode
\includegraphics[width=8.75cm,angle=0]{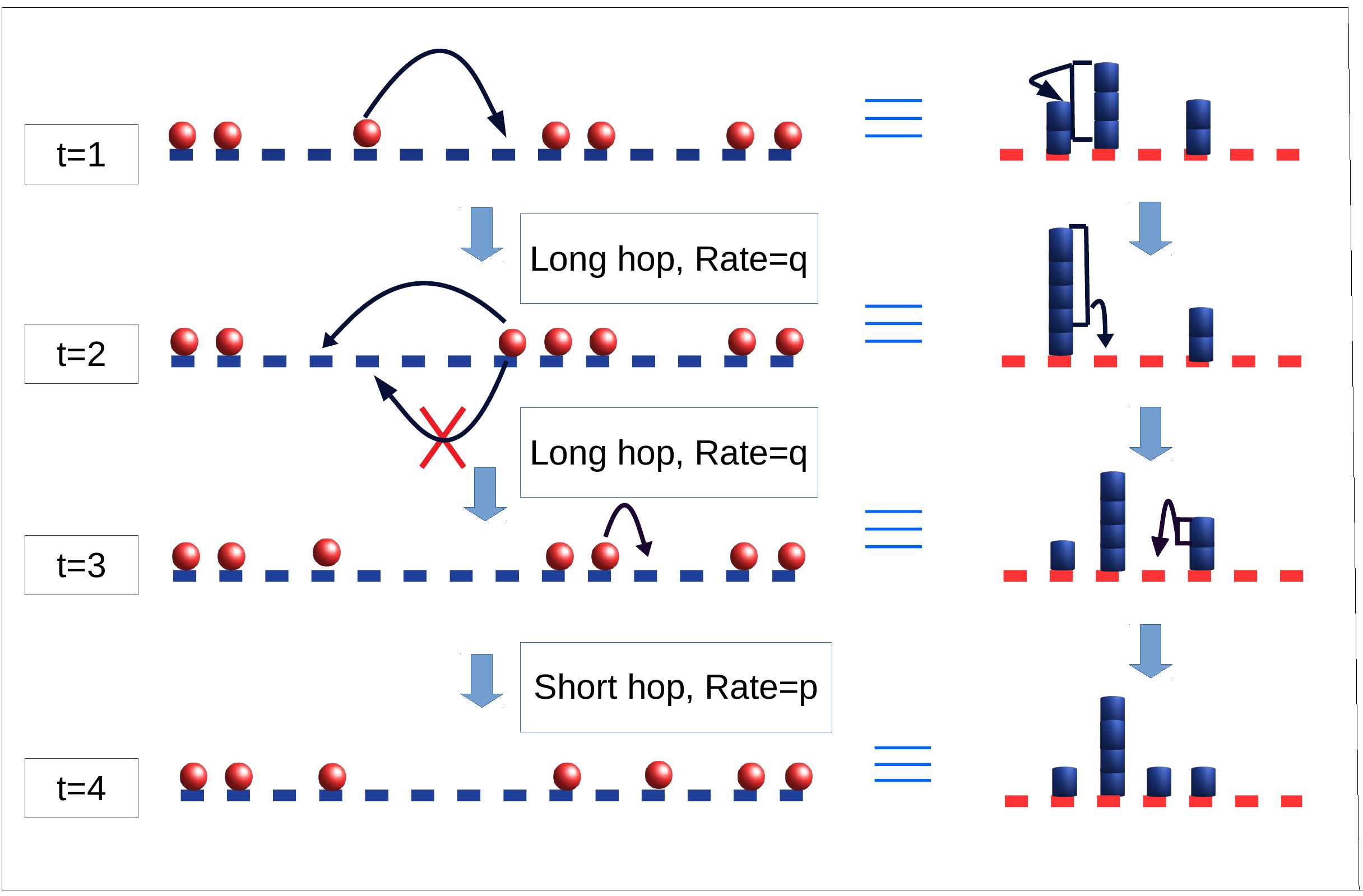}
\caption{ Schematic diagram to illustrate the mapping between gLHM and UgLHM in one dimension in a few successive time-steps; we consider here gLHM with localized hop-length distribution $\phi(l') = \delta_{l',l}$ with $l = 4$.
The filled circles (red) are particles in gLHM and filled blue squares are masses in UgLHM (gaps in gLHM). Maximum possible hop-length in gLHM in this particular case is $l=4$, which, in UgLHM, corresponds to the maximum amount of mass, which can be
transferred at any time. The ``crossed'' arrow indicates the impossibility of the time-reversed hopping process, demonstrating violation of detailed balance in the system.  }
\label{model}
\end{center}
\end{figure}

In this section, we introduce generalized long-hop model (gLHM), which consists of $N$ hardcore particles moving on a one-dimensional periodic lattice of $L$ sites. Due to hardcore constraint, a lattice site can be occupied by at most one particle and crossing between two particles are not allowed. A particle moves according to the following dynamical rules. 
\\
\\
(A) Short range hop: With rate $p$, a particle makes a short range hop of unit length, to its left or right nearest neighbor with equal probability $1/2$, provided that the destination site is vacant. \\
\\
(B) Long range hop: With rate $q$, a particle, say $k$th one, makes a long range hop, to its left or right with equal probability $1/2$, provided there is an empty stretch of vacancies in its hopping direction. At any instant of time, the long-range hop length $l$ is drawn from a probability distribution $\phi(l)$. More specifically, in case of rightward (leftward) hopping, if the gap (i.e., an empty lane consisting of consecutive vacancies) between $k$th and $(k+1)$th [$(k-1)$th] particles is less than $l$, the particle hops to the site adjacent to its nearest occupied site in the hopping direction. However, if the gap in the hopping direction is greater than or equal to $l$, the particle hops the maximum possible distance $l$. We categorize long range hop into two: Finite range hop (FRH) having a typical long-hop length, which is finite, and infinite range hop (IRH) where the typical long-hop length is infinitely large.

We specify a microscopic configuration $\{ \eta_i \}$ by the occupation variable $\eta_i$ at site $i=1, \dots, L$ where $\eta_i=1$ if site $i$ is occupied, otherwise $\eta_i=0$. The total number of particles is conserved and we denote density as $\rho = N/L$. We define a dimensionless parameter $\tilde q=q/(p+q)$, called activity, which parametrizes the competition between short and long hops.  Clearly, for $q=0$ (in the absence of long range hopping), gLHM reduces to the well studied model of simple symmetric exclusion process \cite{SSEP}.

Interestingly, a one-dimensional gLHM with $L$ sites and $N$ particles can be mapped to a one dimensional unbounded model \cite{Hanney}, called here UgLHM, having $N$ ``sites'' and $(L-N)$ ``particles'' and having no hardcore constraint, i.e., the occupation number at a site in UgLHM is unbounded. 
This particular mapping is used later in Sec. \ref{sec-IRH} where we explicitly calculate the transport coefficients for infinite range hopping.
According to the convention we follow here, $k$th particle in gLHM is considered $k$th lattice site in UgLHM and the gap or number of ``holes'' between $k$th and $(k+1)$th particle in gLHM is considered occupancy number or mass at $k$th site in UgLHM. Thus density $\rho$ in gLHM is related to density $\rho'$ in UgLHM as
\begin{equation}
\rho' = \frac{L-N}{N} = \left( \frac{1}{\rho} - 1 \right).
\label{rho-rhop}
\end{equation}  
Accordingly, the dynamical rules in gLHM can be translated to UgLHM as follows. With rate $p$,  a single unit of mass in UgLHM (equivalently, a ``hole'' in gLHM) is chipped off and transferred, to right or left with equal probability $1/2$, to the nearest neighbor site (this particular dynamical rule corresponds to short range hop in gLHM). With rate $q$, two kinds of hopping events are possible: If the mass (number of ``particles'') at a site in UgLHM is greater than $l$, only $l$ unit of mass is fragmented, transferred  to its right or left neighbor with probability $1/2$ and eventually coalesce with the mass at the destination site; otherwise, the whole mass is transferred to its right or left neighbor with probability $1/2$ and coalesce with the mass at the destination site. See the schematic diagram in Fig. \ref{model} for the update rules in both the models and their correspondence. Note that, for generic parameter values, these models violate Kolmogorov criterion and, consequently, detailed balance because some of the hopping events cannot be time-reversed. For example, consider a gLHM with hop-length distribution $\phi(l') = \delta_{l'l}$ with $l = 4$, where the impossibility of a time-reversed path in a particular hopping event is demonstrated in Fig. \ref{model} (indicated by a ``crossed'' arrow). To show the violation of Kolmogorov criterion, one constructs a closed loop in the trajectory space, containing at least one event which cannot be time reversed. Therefore long range hops are responsible for breaking time-reversibility and driving the system out of equilibrium.

\section{Hydrodynamics}

Hydrodynamic time evolution provides large-scale spatio-temporal behaviors of slow variables in a system. Since the total number of particles are conserved in gLHM, the slow variable in our case is the local particle-number density $\rho(x,t)$ at position $x$ and time $t$. Our aim in this paper is to obtain a large-scale hydrodynamic structure of the time-evolution of density field $\rho(x, t)$, which is governed by a continuity equation, 
\begin{equation}
\label{continuity}
\frac{\partial \rho(x,\tau)}{\partial \tau} = -\frac{\partial}{\partial x} \left[ -D(\rho)\frac{\partial \rho}{\partial x} + \chi(\rho) F \right] \equiv - \frac{\partial J}{\partial x},
\end{equation}
through a constitutive relation between local density $\rho(x,t)$ and hydrodynamic current $J(\rho) = -D(\rho) {\partial \rho}/{\partial x} + \chi(\rho) F$, defined using two transport coefficients - the bulk-diffusion coefficient $D(\rho)$ and the conductivity $\chi(\rho)$. The first term in the current arises according to Fick's law where a nonuniform density profile contributes to a diffusive current $J_D(\rho(x,t)) = - D(\rho) \partial \rho(x,t)/\partial x$ and the second term in the current provides a drift current $J_d = \chi(\rho) F$, which is essentially the (linear) response to a small perturbation due to an externally applied biasing force of magnitude $F$.

To calculate the conductivity $\chi(\rho)$ in the presence of a small biasing force $F$ (say, counter-clockwise along the ring), we modify, by following macroscopic fluctuation theory  \cite{Bertini_PRL2001}, the original (unbiased) hopping rate $c_{i \rightarrow j}$ from site $i$ to $j$ to a biased hopping rate  
\begin{align}
c_{i \rightarrow j}^F=c_{i \rightarrow j} \exp \left( \frac{\Delta e_{ij}}{2} \right) \simeq c_{i \rightarrow j} \left[ 1 + \frac{F(j-i)a}{2} \right],
\end{align}
which  is linearized in the limit of small force $F$, with $\Delta e_{ij}={\Delta m_{i \rightarrow j} F(j-i)a/2}$ being an ``energy cost'' for transferring $\Delta m_{i \rightarrow j}$ number of particles  from site $i$ to $j$; for gLHM, $\Delta m_{i \rightarrow j}=1$, which is the number of particle transferred at a time, and $a=1$ the lattice spacing. 
In gLHM with modified hopping rates, each particle hops with rates, which are slightly larger in the direction of the applied force than that in the opposite direction.

Let us start with the simplest case of gLHM with a localized distribution $\phi(l')=\delta_{l'l}$ of long-hop lengths, i.e., long-hop with a fixed hop-length $l$. We denote separately the modified (or biased) hop rates in each direction: The modified long-hop rates as $q^F_R(l)$ and $q^F_L(l)$  and similarly the modified short-hop rates as $p^F_R$ and $p^F_L$, where the subscripts ``R'' and ``L'' denote  anti-clockwise (in the direction of the biasing force) and clock-wise (opposite to biasing force) hopping directions of particles, respectively.  To calculate the rate of change of average occupancy $\rho_i = \langle \eta_i (t) \rangle$ or local density at site $i$, we consider all possible ways of gaining and loosing a particle at site $i$. Clearly there are total four contributions at a site $i$, two of which are associated with loss of a particle, i.e., outward fluxes $J_R^{-}(i)$ towards right and $J_L^{-}(i)$ towards left and the remaining two with gain of a particle, i.e., inward fluxes $J_R^{+}(i)$ towards right and $J_L^{+}(i)$ towards left. Now the rate of change of average occupation $\langle \eta_i (t) \rangle$ can be written as 
\begin{equation}
\label{C1}
\frac{\partial \langle \eta_i (t) \rangle}{\partial t} = \frac{\partial \rho_i (t)}{\partial t} = J^{+}_R(i) + J^{+}_L(i) - J^{-}_R(i) - J^{-}_L(i).
\end{equation} 
Let us explicitly consider the hopping events of a particle at site $i$ toward its right direction, corresponding to the term $J^{-}_R(i)$ in Eq. (\ref{C1}), which has two contributions: the flux contribution $J^{-}_{R,sh}(i)$ due to short-hop and the other one $J^{-}_{R,l}(i)$ due to long-hop, as shown below.

{\it Short-hop contribution.$-$} For short-range hop, a particle hops by a unit distance, provided the destination site is empty. Therefore, $i$th site can gain particle from the nearest neighbor $(i+1)$ or $(i-1)$; on the other hand, site $i$ can lose a particle when the particle hops to the nearest neighbor $(i+1)$ or $(i-1)$. Therefore, the corresponding loss rate for a particle, moving to right from site $i$ to $(i+1)$, is given by  
 \begin{eqnarray}
 J^{-}_{R,sh}(i) &=& \frac{1}{2} p^{F}_R \langle \eta_{i}\overline{\eta}_{i+1} \rangle 
 \nonumber \\
 &=& \frac{p}{2} \left( 1+\frac{Fa}{2} \right) \langle \eta_{i} (1-\eta_{i+1})\rangle + {\cal O}(F^2),
 \nonumber
 \end{eqnarray}
where angular brackets denote steady-state averages.

{\it Long hop contribution.$-$}  In this case, depending on the number of consecutive vacant sites, or gap size $g$, two different kinds of hopping events are possible from site $i$ as following. 
\\
\\
{\it Case I.$-$} If $g < l$, the particle at site $i$ can hop only by length $g$  as the nearest occupied site is located at $(i+g+1)$. Therefore the corresponding loss rate is given by
\begin{eqnarray}
J^{-}_{R,<} (i\rightarrow{i+g}) &=& \frac{1}{2} q^F_R(g) \langle \eta_{i}\overline{\eta}_{i+1}\overline{\eta}_{i+2} \dots \overline{\eta}_{i+g}\eta_{i+g+1} \rangle \nonumber.
\end{eqnarray}
\\
\\
{\it Case II.$-$} If $g \geq l$, the particle hops by maximum possible length $l$ and then resides at $(i+l)$th site and the corresponding loss rate is given by
 \begin{eqnarray}
J^{-}_{R,\ge}(i \rightarrow i+l) &=& \frac{1}{2}q^F_R(l)\langle \eta_{i} \overline{\eta}_{i+1} \overline{\eta}_{i+2} \dots \overline{\eta}_{i+l}\rangle.
\nonumber
\end{eqnarray}
The total loss rate corresponding to rightward outflux of particle from site $i$, considering all possible gap sizes, can be written as
\begin{eqnarray} \nonumber
J_{R,l}^{-}(i)=\sum_{g=1}^{l-1}J^{-}_{R,<} (i\rightarrow{i+g}) + J^{-}_{R,\ge}(i \rightarrow i+l).
\end{eqnarray}
Now by denoting the correlation functions as
\begin{eqnarray}\label{Al}
{\cal A}^{(l)}_i = \langle \overline{\eta}_{i-l+1} \dots \overline{\eta}_{i-1} \overline{\eta}_{i}\rangle,
\\ \label{Bl}
{\cal B}^{(l+2)}_i = \langle \eta_{i-l-1} \overline{\eta}_{i-l} \overline{\eta}_{i-l+1} \dots \overline{\eta}_{i-1} \eta_i \rangle,
\\ \label{B2}
{\cal B}^{(2)}_i = \langle \eta_{i-1} \eta_i \rangle ,
\end{eqnarray}
we write various rightward fluxes in linear order ${\cal O}(F)$ of the biasing force $F$, as given below,
\begin{eqnarray}
J^{-}_{R,sh}(i) &=& \frac{p}{2} \left(1 + \frac{Fa}{2} \right) (\rho_{i}-\mathcal{B}_{i+1}^{(2)}) + {\cal O}(F^2),
\nonumber \\
J^{-}_{R,<} (i\rightarrow{i+g}) &=& \frac{q}{2}\left(1+ \frac{Fga}{2} \right) \mathcal{B}_{i+g+1}^{(g+2)} + {\cal O}(F^2), 
\nonumber \\
J^{-}_{R,\ge}(i \rightarrow i+l) &=& \frac{q}{2} \left(1 +   \frac{Fla}{2} \right) \left( \mathcal{A}_{i+l}^{(l)} - \mathcal{A}_{i+l}^{(l+1)} \right)  + {\cal O}(F^2)
\nonumber
\end{eqnarray}
The net loss rate $J_{R}^{-}(i)$ due to both short range and long range hop is given by
\begin{eqnarray}
\label{loss_current}
J_{R}^{-}(i)=J^{-}_{R,sh}(i)+ J_{R,l}^{-}(i).  
\end{eqnarray}
 We can calculate other loss and gain rates in a similar way; for detailed calculations of $J_{L}^{-}(i)$, $J_{R}^{+}(i)$ and $J_{L}^{+}(i)$, see the Appendix Sec. A.

\subsection{Finite range hopping}
\label{sec-frh}

In this section, we set up the continuity equation for the density field for gLHM with finite range hopping, i.e., the long-hop length is chosen from a distribution $\phi(l')=\delta_{l',l}$ with finite long-hop range $l$. As derived in Eq. (\ref{loss_current}), the rightward loss rate from site $i$ is given by, upto linear order ${\cal O}(F)$ of force,
\bea
J_{R}^{-}(i) \simeq  \frac{p}{2} (1 + \frac{Fa}{2} )(\rho_{i} - \mathcal{B}_{i+1}^{(2)})
 + \sum_{g=1}^{l-1} \frac{q}{2} (1+\frac{Fga}{2}) \mathcal{B}_{i+g+1}^{(g+2)} 
\nonumber \\  
+ \frac{q}{2} (1 + \frac{Fla}{2}) \left( \mathcal{A}_{i+l}^{(l)} - \mathcal{A}_{i+l}^{(l+1)} \right).~
\label{J1}
\eea
Similarly, as shown in the Appendix Sec. A, we write the leftward gain rate,
\bea
J_{L}^{+}(i) \simeq \frac{p}{2} ( 1-\frac{Fa}{2} )(\rho_{i+1}-\mathcal{B}_{i+1}^{(2)}) + \sum_{g=1}^{l-1}\frac{q}{2}(1-\frac{Fga}{2}) \mathcal{B}_{i+g}^{(g+2)} 
\nonumber \\ 
+ \frac{q}{2} \left( 1 - \frac{Fla}{2} \right) \left(\mathcal{A}_{i+l-1}^{(l)}-\mathcal{A}_{i+l}^{(l+1)} \right), ~
\label{J2}
\eea
the rightward gain rate,
\bea
J_{R}^{+}(i) \simeq \frac{p}{2}(1+\frac{Fa}{2})(\rho_{i-1}-\mathcal{B}_{i}^{(2)})+ \sum_{g=1}^{l-1}\frac{q}{2}(1+\frac{Fga}{2}) \mathcal{B}_{i+1}^{(g+2)} 
\nonumber \\
 + \frac{q}{2}(1+\frac{Fla}{2})\left(\mathcal{A}_{i}^{(l)}-\mathcal{A}_{i}^{(l+1)} \right),~
\label{J3}
\eea
and the leftward loss rate, 
\bea
J_{L}^{-}(i) \simeq \frac{p}{2}(1-\frac{Fa}{2})(\rho_{i}-\mathcal{B}_{i}^{(2)})+ \sum_{g=1}^{l-1}\frac{q}{2}(1-\frac{Fga}{2}) \mathcal{B}_{i}^{(g+2)} 
\nonumber \\ 
+ \frac{q}{2}(1-\frac{Fla}{2})\left(\mathcal{A}_{i-1}^{(l)}-\mathcal{A}_{i}^{(l)}\right).~
\label{J4}
\eea
Substituting all loss and gain rates from Eqs. (\ref{J1}),  (\ref{J2}), (\ref{J3}) and (\ref{J4}) into Eq. (\ref{C1}), we obtain the time-evolution of local density, which, in the leading order ${\cal O}(F)$ of the biasing force $F$, is recast below in a somewhat long, but an interesting form,
\begin{widetext}
\bea 
\frac{\partial \rho_{i}}{\partial t} \simeq \frac{q}{2} \left[ \{ {\cal A}^{(l)}_{i+l-1} - {\cal A}^{(l)}_{i+l} \} - \{ {\cal A}^{(l)}_{i-1} - {\cal A}^{(l)}_{i} \} \right] 
- \frac{qFl}{4} \left[ \{ {\cal A}^{(l)}_{i+l-1} + {\cal A}^{(l)}_{i+l} \} 
- \{ {\cal A}^{(l)}_{i-1} + {\cal A}^{(l)}_{i} \} - 2 \{ {\cal A}^{(l+1)}_{i+l} - {\cal A}^{(l+1)}_{i} \} \right]
 \nonumber \\ 
+ \sum_{g=1}^{l-1} \frac{q}{2} \left[ \{ {\cal B}^{(g+2)}_{i+1} - {\cal B}^{(g+2)}_i \} - \{ {\cal B}^{(g+2)}_{i+g+1} - {\cal B}^{(g+2)}_{i+g} \} \right] 
- \sum_{g=1}^{l-1} \frac{qFg}{4} \left[ \{ {\cal B}^{(l+2)}_{i+g+1} + {\cal B}^{g+2}_{i+g} \} - \{ {\cal B}^{(g+2)}_{i+1} + {\cal B}^{(g+2)}_i \} \right]
\nonumber \\
+ \frac{p}{2} \left[ \rho_{i+1} - 2 \rho_i + \rho_{i-1} \right] + \frac{pF}{4} \left[ \{ \rho_{i-1} - \rho_{i+1} \} + 2 \{ {\cal B}^{(2)}_{i+1} - {\cal B}^{(2)}_i \} \right],~~
\label{C2}
\eea
\end{widetext}
where quantities inside the curly brackets are written in the form of a gradient of observables, leading to a continuity equation for local density as follows. Taking diffusive scaling limit $i \rightarrow x=i/L$, $t \rightarrow t/L^2$ and $a \rightarrow 1/L$, where the observables are assumed to vary slowly in space and time and therefore to take {\it local steady-state} values \cite{Bertini_PRL2001, Landim}, we expand ${\cal A}^{(l)}_i(t) \equiv {\cal A}^{(l)}[\rho(x,t)]$ in Taylor's series around local density $ \rho_i(t) \equiv \rho(x, t)$ and obtain, upto ${\cal O}[(1/L)^2]$,
\bea 
{\cal A}^{(l)}_{i+l} \simeq {\cal A}^{(l)}[\rho(x,t)] + \frac{l}{L} \frac{\partial {\cal A}^{(l)}[\rho(x,t)]}{\partial x} 
\nonumber \\
+ \frac{l^2}{2 L^2} \frac{\partial^2 {\cal A}^{(l)}[\rho(x,t)]}{\partial x^2},
\nonumber
\eea
and similarly for ${\cal B}^{(l+2)}_{i+l}$ and ${\cal B}^{(2)}_{i+1}$, etc. Using the above Taylor series expansion in Eq. (\ref{C2}) and collecting terms upto ${\cal O}(1/L^2)$, we obtain in the diffusive scaling limit the desired hydrodynamics of gLHM as a continuity equation $\partial \rho(x,t)/\partial t + \partial J(\rho(x,t))/\partial x=0$ for local density $\rho(x,t)$,
\be 
\frac{\partial \rho(x,t)}{\partial t} = - \frac{\partial}{\partial x} \left[ - D_l(\rho) \frac{\partial \rho}{\partial x} + \chi_l(\rho) F \right],
\label{Hydro-l}
\ee
where the two density-dependent transport coefficients - the bulk-diffusion coefficient and the conductivity are given by 
\bea  
D_l(\rho) =  \frac{p}{2} - \frac{q}{2} \sum_{l'=1}^{l-1} l' \frac{\partial {\cal B}^{(l'+2)}(\rho)}{\partial \rho } - \frac{ql}{2} \frac{\partial {\cal A}^{(l)}(\rho)}{\partial \rho},
\label{Dl}
\\
\chi_l(\rho) = \frac{1}{2} \left[ q \sum_{l'=1}^{l-1} l'^{2} {\cal B}^{(l'+2)}(\rho) + p {\cal B}^{(2)}(\rho) \right] ~~~~~~~~
\nonumber \\
~~~~~~~~~~~~~~~~~~+ \frac{ql^2}{2} \left[ {\cal A}^{(l)}(\rho) - {\cal A}^{(l+1)}(\rho) \right],
\label{chi-l}
\eea
respectively; for details, see the Appendix Sec. A.4. In deriving the above hydrodynamic evolution of density field, we have  essentially established the constitutive relation between local current $J(\rho)$ and local density $\rho(x,t)$ as $J(\rho) = J_D(\rho) + J_d(\rho)$ where total current is split into two parts - the diffusive current $J_D(\rho) = - D_l(\rho) \partial \rho(x,t)/\partial x$ and the drift current $J_d(\rho) = \chi_l(\rho) F$. The above Eqs. (\ref{Dl}) and (\ref{chi-l}) constitute the first main results  of the paper. For a general distribution $\phi(l')$ of long-hop length $l'$, the bulk-diffusion coefficient $D(\rho)$ and the conductivity $\chi(\rho)$ is obtained by performing a weighted sum of Eqs. (\ref{Dl}) and (\ref{chi-l}) over all possible hop-lengths: $D(\rho) = \sum_{l'} \phi(l') D_{l'}(\rho)$ and $\chi(\rho) = \sum_{l'} \phi(l') \chi_{l'}(\rho)$; generalizations of the results to higher dimensions is straightforward.  Although, at this stage, we do not have explicit expressions of the transport coefficients, one however can readily calculate them numerically as a function of density and can verify the above hydrodynamic structure Eq. (\ref{Hydro-l}). In the next section, we study the interesting special case of gLHM with infinite range hopping, which is analytically tractable, exhibits phase transition and for which one can calculate the two transport coefficients explicitly as a function of density.

\subsection{Infinite range hopping}
\label{sec-IRH}

For finite range hopping, the transport coefficients as in Eqs. (\ref{Dl}) and (\ref{chi-l}) remain finite and there is no phase transition as such. However, the situation changes when the typical length-scale in the long range hop diverges. To demonstrate this point, we consider a special case of the infinite range hopping where the hop-length distribution has the following form: $\phi(l')= \delta_{l' l}$ with $l \rightarrow \infty$.

The dynamics for short range hopping is exactly the same as described in the previous section. However, due to infinite range hopping, the dynamics for long range hop is slightly modified. Now, during a long range hop, a particle at a site $i$ always hops the maximum possible distance along an empty lane in its hopping direction (which is still chosen symmetrically with probability $1/2$). That is, the particle at site $i$ hops a distance $g$ - the size of the gap in front of it.

We outline below the calculation techniques; for details, see the Appendix Sec. B. First of all, in the case of infinite range hopping with $l \rightarrow \infty$, some simplifications occur as the terms involving ${\cal A}^{(l)}_i$'s drop out from Eqs. (\ref{Dl}) and (\ref{chi-l}), leading to the bulk-diffusion coefficient and the conductivity $\chi(\rho)$ as given below,
\bea
D(\rho) = \frac{p}{2} -\frac{q}{2} \sum_{l'=1}^{\infty} l' \frac{\partial {\cal B}^{l'+2}(\rho)}{\partial \rho},
\nonumber \\
\chi(\rho) = \frac{p}{2} {\cal B}^{(2)}(\rho) + \frac{q}{2} \sum_{l'=1}^{\infty} l'^{2} {\cal B}^{l'+2}(\rho).
\nonumber
\eea
But we still have to determine ${\cal B}^{l'+2}(\rho)$ and ${\cal B}^{(2)}(\rho)$  as a function of $\rho$. To this end, we exploit the previously described mapping between gLHM and its {\it unbounded} version - UgLHM, for which the infinite range hopping translates into the diffusion of the individual masses as a whole, thus incorporating complete aggregation of neighboring masses in UgLHM. Similar versions of UgLHM have been studied in the past in the context of mass aggregation and gelation processes \cite{Barma_PRL1998, Krapivsky_PRE1996, MAM}. However, the large-scale hydrodynamic structure of these mass-aggregating systems is still largely unexplored. Below we focus on hydrodynamics of gLHM with infinite range hopping; hydrodynamics of the corresponding unbounded version of the model (i.e., UgLHM with aggregation) will be presented elsewhere \cite{unbounded}.

We now proceed by noting that the mass or the gap distribution $P(g_k|\rho')$ at `site' $k$ in UgLHM with a given $\rho'$ is related to the required correlations in gLHM as 
$$
{\cal B}^{l+2}(\rho) = \rho P(g=l|\rho'),
$$ and 
$$
{\cal B}^{(2)}(\rho) = \rho P(g=0|\rho') = \rho [1- {\cal c}(\rho')],
$$ 
where ${\cal c}(\rho')$ is the occupation probability in UgLHM. Using the identities
$$
\rho' = \langle g \rangle = \sum_g g P(g|\rho'),
$$
and
$$
\sum_{l'=1}^{\infty} l' \frac{\partial {\cal B}^{l'+2}(\rho)}{\partial \rho} = \frac{\partial (\rho \rho')}{\partial \rho},
$$ 
we find that one actually requires only the second moment $\theta_2(\rho') = \sum_g g^2 P(g|\rho')$ of the gap distribution $P(g|\rho')$ to obtain the bulk-diffusion coefficient and the conductivity 
\be
D(\rho) = \frac{p}{2} - \frac{q}{2} \frac{d(\rho \rho')}{d\rho},
\ee 
and 
\be
\chi(\rho) = \frac{p}{2} \rho {\cal c}(\rho') + \frac{q}{2} \rho \theta_2(\rho'),
\ee 
respectively, where we ${\cal c}(\rho')$ is the probability that a site is occupied in UgLHM. Now one can immediately calculate $\theta_2(\rho')$ by assuming a statistical independence between neighboring masses in UgLHM, which, as our finite-size scaling analysis indicates, could actually be exact. 
Finally, some further algebraic manipulations, using ${\cal c}(\rho')=\rho'(p - q \rho')/p(1+\rho')$, $\theta_2(\rho')=p \rho'[1+{\cal c}(\rho')]/[p\{1- {\cal c}(\rho')-2 q \rho'\}]$ and $\rho'=1/\rho-1$, give explicit expressions of the two transport coefficients,
\bea
D(\rho) &=& \frac{p+q}{2}
\label{D-inf} 
\\ 
\chi(\rho) &=& \frac{\rho(1-\rho)[(p+q) \rho^2 - 2q \rho + q]}{2[\rho^2-q/(p+q)]},
\label{chi-inf}
\eea
the second main results of the paper; for calculation details, see the Appendix Sec. C. Therefore in gLHM with infinite range hopping, though the bulk-diffusion coefficient remains finite (constant), interestingly, upon approaching a critical density (or activity), the conductivity develops a singularity, a first-order pole, signifying a phase transition beyond a critical density $\rho_c(q)={\tilde q}^{1/2}$ [or a critical activity $\tilde q_c(\rho)=\rho^2$]. As discussed later in Sec. \ref{sec-gnf}, at criticality and beyond (i.e., for $\rho \le \rho_c$ or $\tilde q \ge \tilde q_c$), the bulk of the system behaves like a ``superfluid'', having diverging conductivity $\chi(\rho) \sim \theta_2(\rho) \sim (\rho-\rho_c)^{-1}$ [equivalently, vanishing resistivity $\sim (\rho-\rho_c)$], a direct consequence of diverging gap or vacancy fluctuations, through which cooperativity emerges in the system.

\subsection{Density Relaxation}
\label{sec-den-rel}

\begin{figure}[tb!]
\centering
\subfigure
{\includegraphics[width=1.0 \linewidth]{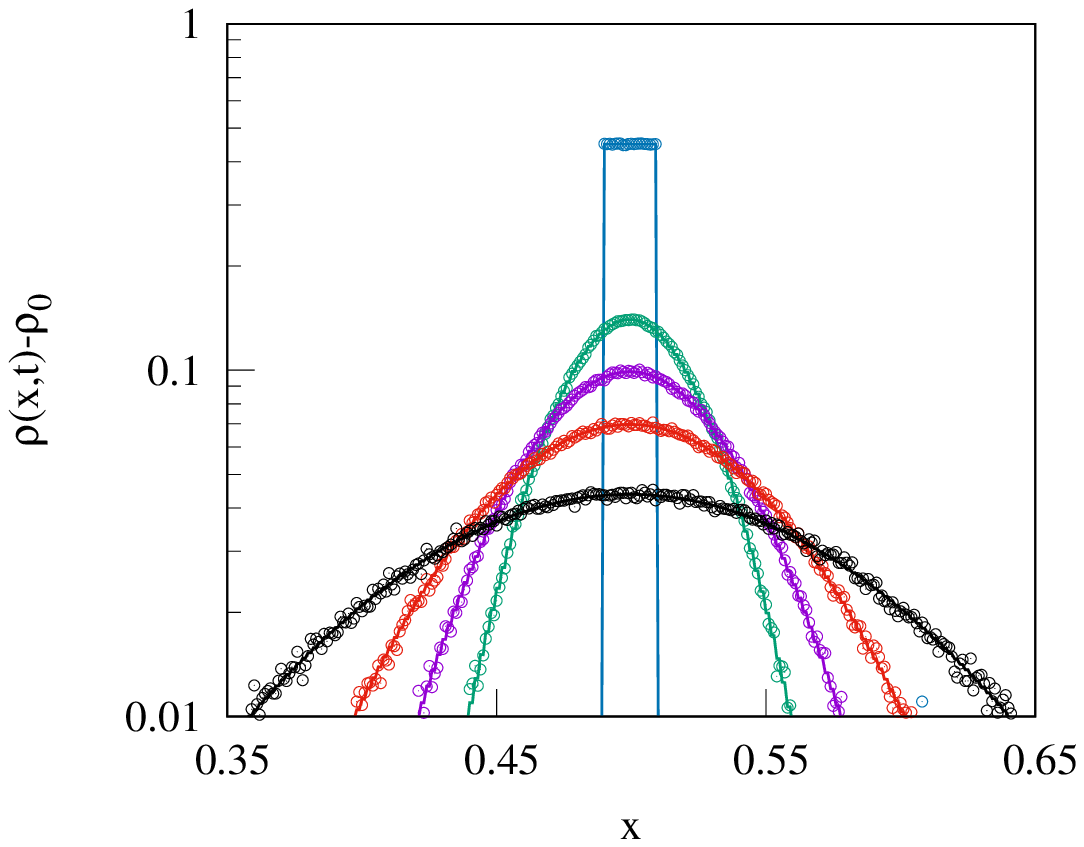}}\hfill
\subfigure
{\includegraphics[width=1.0 \linewidth]{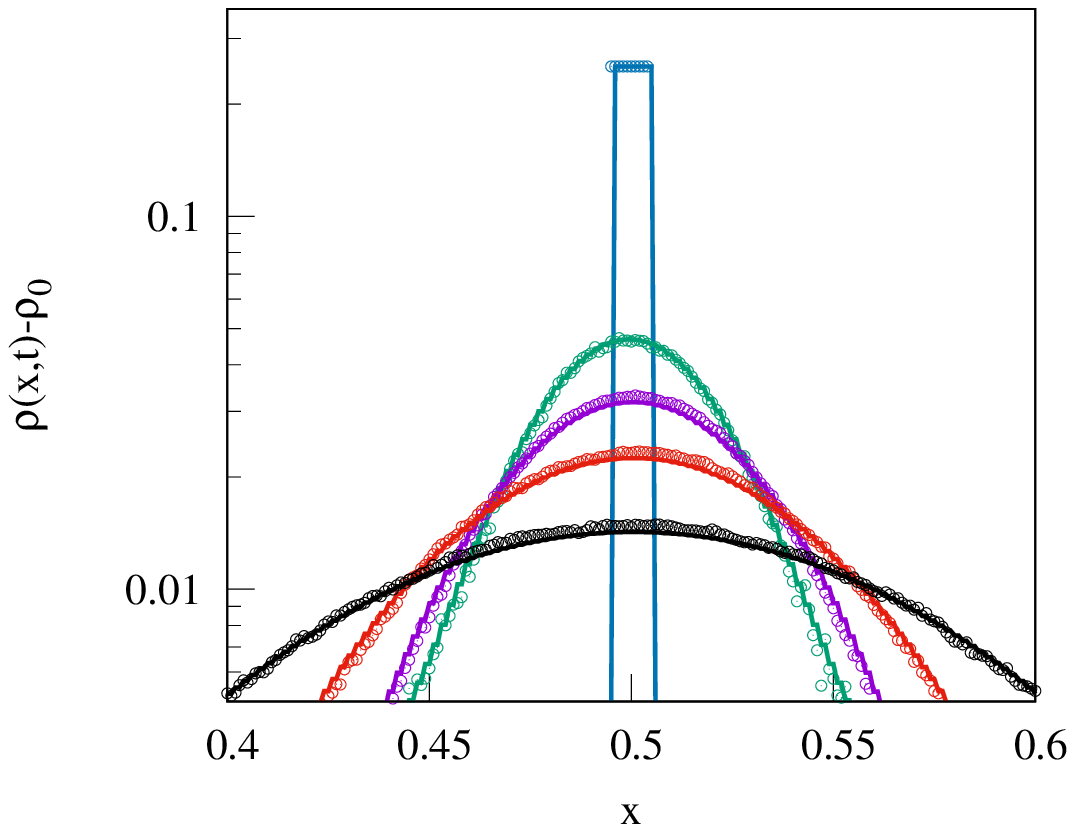}}
\caption{Density relaxation in gLHM from two-step initial condition. Density profiles $\delta \rho(x,t) = \rho(x,t) - \rho_0$ (line) obtained by numerically integrating Eq. (\ref{continuity-F0}) are compared with that obtained from microscopic simulations (points) at $t=0$ (blue points, initial profile), $t=0.5 \times 10^{-3}$ (green), $10^{-3}$ (magenta), $2 \times 10^{-3}$ (red) and $5 \times 10^{-3}$ (black). Top panel: Finite range hopping with $l=2$, $\rho_0=0.5$ and $\rho_1=0.5$; bottom panel: infinite range hopping with $l \rightarrow \infty$, $\rho_0=0.75$, $\rho_1=0.25$. We take $w=0.1$ and $L=1000$ in both cases. }
\label{den-rel}
\end{figure}

In this section, we study density relaxation from an initial density-perturbation in the generalized long-hop model, with the original unbiased hopping rates ($F=0$). As derived in the previous sections, the process of density relaxation is governed by Eq. (\ref{continuity}) with $F=0$,
\begin{equation}
\frac{\partial \rho(x,t)}{\partial t}=\frac{\partial}{\partial x}\left(D(\rho)\frac{\partial \rho(x,t)}{\partial x}\right),
\label{continuity-F0}
\end{equation}
where $x=i/L$ is rescaled position and $t$ is hydrodynamic rescaled time (in unit of $L^2$). To verify the above hydrodynamic time-evolution of density field, we solve Eq. (\ref{continuity-F0}), with a suitable initial condition $\rho(x, t=0)$, by performing numerical integration of the equation for finite range hopping with $l=2$ as well as infinite range hopping ($l \rightarrow \infty$). From Eq. (\ref{Dl}), the bulk-diffusion coefficient for $l=2$ can be written as  
\begin{eqnarray}\label{D l2}
D (\rho)&=&  \frac{p}{2}-\frac{q}{2}\frac{\partial {\cal B}^{(3)}(\rho)}{\partial \rho} - q \frac{\partial {\cal A}^{(2)}(\rho)}{\partial \rho} \nonumber \\ &=& \frac{p}{2}-\frac{q}{2}\frac{\partial [\rho P(g=1|\rho)]}{\partial \rho}-q\frac{\partial P(g \geq 2|\rho)}{\partial \rho},
\end{eqnarray}
where ${\cal B}^{(3)} = \rho P(g=1|\rho)$ with $P(g=1|\rho)$ being the probability of a gap of unit size and ${\cal A}^{(2)} = P(g \geq 2|\rho)$ is the probability of a gap of size greater than or equal to $2$, provided density being $\rho$. As we do not have the explicit expressions for the probabilities $P(g=1|\rho)$ and $P(g \ge 2|\rho)$, we obtain, the numerical values of bulk-diffusion coefficient as a function of density by directly calculating the above probabilities from simulations. For infinite range hopping, we use the bulk-diffusion coefficient as given in Eq. (\ref{D-inf}) to solve Eq. (\ref{continuity-F0}). We take the initial density perturbation as a two-step function of height $\rho_1$ and width $w$ over a uniform density profile $\rho_0$, i.e., the initial density profile is given by  
\bea \phi(x) =  \left\{
\begin{array} {cc}
   \rho_0 + \rho_1 &\rm{~~~for~|x-\frac{1}{2}| < \frac{w}{2},}  \cr
   \rho_0 & \rm{~~~otherwise.}
\end{array} \right. \nonumber
\eea 
In Fig. \ref{den-rel}, we plot density profile $\delta \rho(x,t) = \rho(x,t) - \rho_0$, obtained from simulations, as a function of rescaled position $x=i/L$ at various hydrodynamic times $t=0$ (blue points, initial profile), $t=0.5 \times 10^{-3}$ (green), $10^{-3}$ (magenta), $2 \times 10^{-3}$ (red) and $5 \times 10^{-3}$ (black) for finite range hopping ($l=2$, $\rho_0=0.5$, top panel) and infinite range hopping ($l \rightarrow \infty$, $\rho_0 = 0.75$, bottom panel); in both cases, we take $L=1000$ and $p=q=1/2$. The simulations (points) compare quite well with the hydrodynamic theory (lines).

\subsection{Einstein Relation}
\label{sec-Ein-Rel}

In this section, we demonstrate, using macroscopic fluctuation theory, how the hydrodynamic transport coefficients can be related to particle-number fluctuations in generalized long-hop model. While hydrodynamics provides average behavior of the system on a local coarse-grained level, there are also fluctuations in local observables, such as density $\hat \rho(x,t)$ and current $\hat j(x,t)$, whose probabilities are provided by macroscopic fluctuation theory in terms of the two transport coefficients \cite{Bertini_PRL2001}.  More specifically, if the transport coefficients are known as a function of density, macroscopic fluctuation theory predicts the steady-state joint probability ${\cal P}[\{ \hat \rho(x,t), \hat j(x,t) \}]$ of density $\hat \rho(x,t)$ and current $\hat j(x,t)$ trajectories  in a given domain of space $x \in \Lambda$ and time $t \in [0,T]$ \cite{Bertini_PRL2001, Derrida}, 
\bea
{\cal P}[\{ \hat \rho(x,t), \hat j(x,t) \}] \sim \int {\cal D} \hat \rho  \int {\cal D} \hat j \delta(\partial \hat \rho/\partial t + \partial \hat j/\partial x) 
\nonumber \\
\times \exp \left[ - L \int_0^T dt \int_{\Lambda} dx \frac{ \left\{ j - D(\hat \rho) \partial \hat \rho/\partial x \right\}^2}{4 \chi(\hat \rho)} \right],
\label{MFT}
\eea
where Dirac-delta function imposes the constraint of continuity equation $\partial \hat \rho/\partial t + \partial \hat j/\partial x=0$ and $L$ is the system size. On a periodic domain, the large-deviation probability ${\cal P}[\{ \hat \rho(x) \}]$ of a given density profile $\hat \rho(x,t)$ can be obtained simply in terms of a local equilibrium-like free energy-density functional $V[\hat \rho(x)]$,
\be 
{\cal P}[\{ \hat \rho(x) \}] \sim e^{-L V[\hat \rho(x)]},
\nonumber
\ee
where $V[\hat \rho(x)] = \int_{\Lambda} dx [f(\hat \rho)-f(\rho) - \mu(\rho)(\hat \rho-\rho)]$, $\rho$ is global density, $f(\rho)$ is a nonequilibrium free energy-density function and $\mu(\rho)=df/d\rho$ is a nonequilibrium chemical potential \cite{Bertini_PRL2001}. The functional dependence of free energy density on number density  is determined by integrating a fluctuation-response relation between a nonequilibrium compressibility and number-fluctuation \cite{Das},
\be 
\frac{d \rho}{d \mu} = \sigma^2(\rho),
\label{FR}
\ee 
through an Einstein relation (ER) \cite{Bertini_PRL2001},
\be 
\sigma^2(\rho) = \frac{\chi(\rho)}{D(\rho)},
\label{ER1}
\ee
where $\sigma^2(\rho)$ is the scaled particle-number fluctuation and $\rho$ is the global density. In the above equation, we have defined the scaled number fluctuation as
\be
\sigma^{2}(\rho)=\lim_{l_{sub} \rightarrow \infty} \frac{\langle (\Delta n)^2 \rangle}{l_{sub}}
\nonumber
\ee
where $\langle (\Delta n)^2 \rangle = \langle (n-\langle n \rangle)^2 \rangle$ is the variance of particle-number $n=\sum_{i=1}^{l_{sub}} \eta_i$  in a subsystem of size $l_{sub}$.

\begin{figure}[tb]
\centering
\subfigure
{\includegraphics[width=1.0 \linewidth]{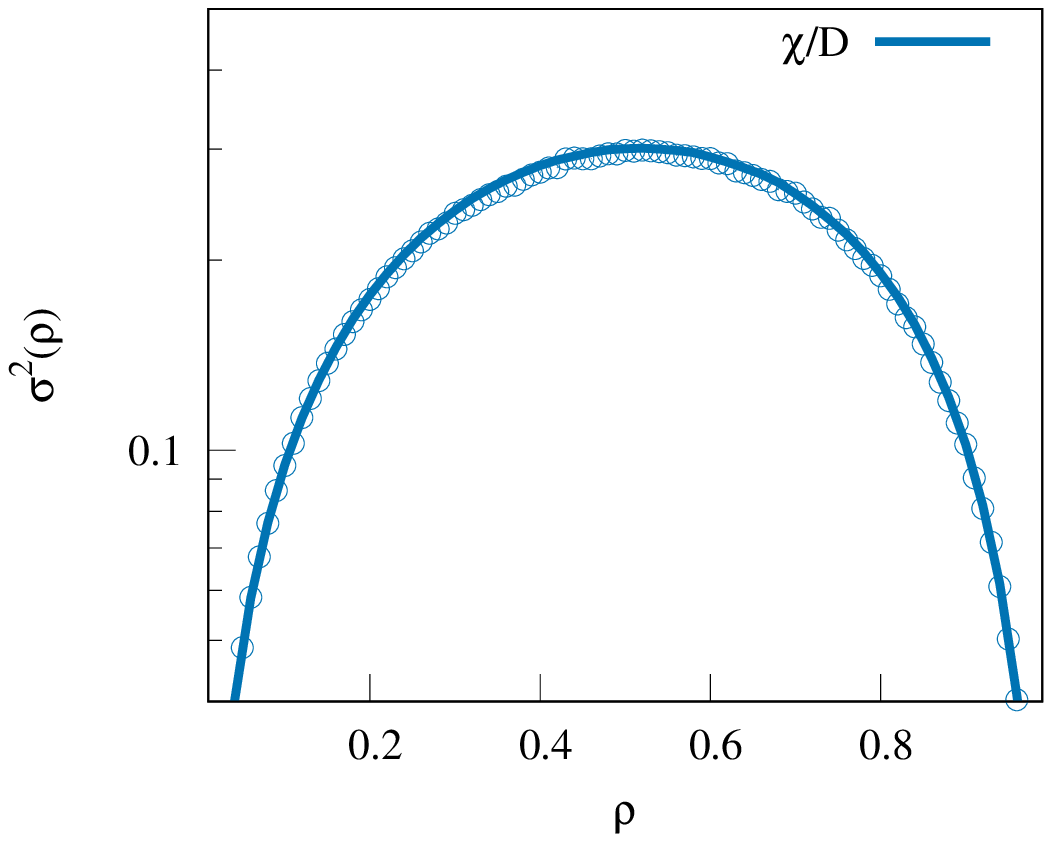}}\hfill
\subfigure
{\includegraphics[width=1.0 \linewidth]{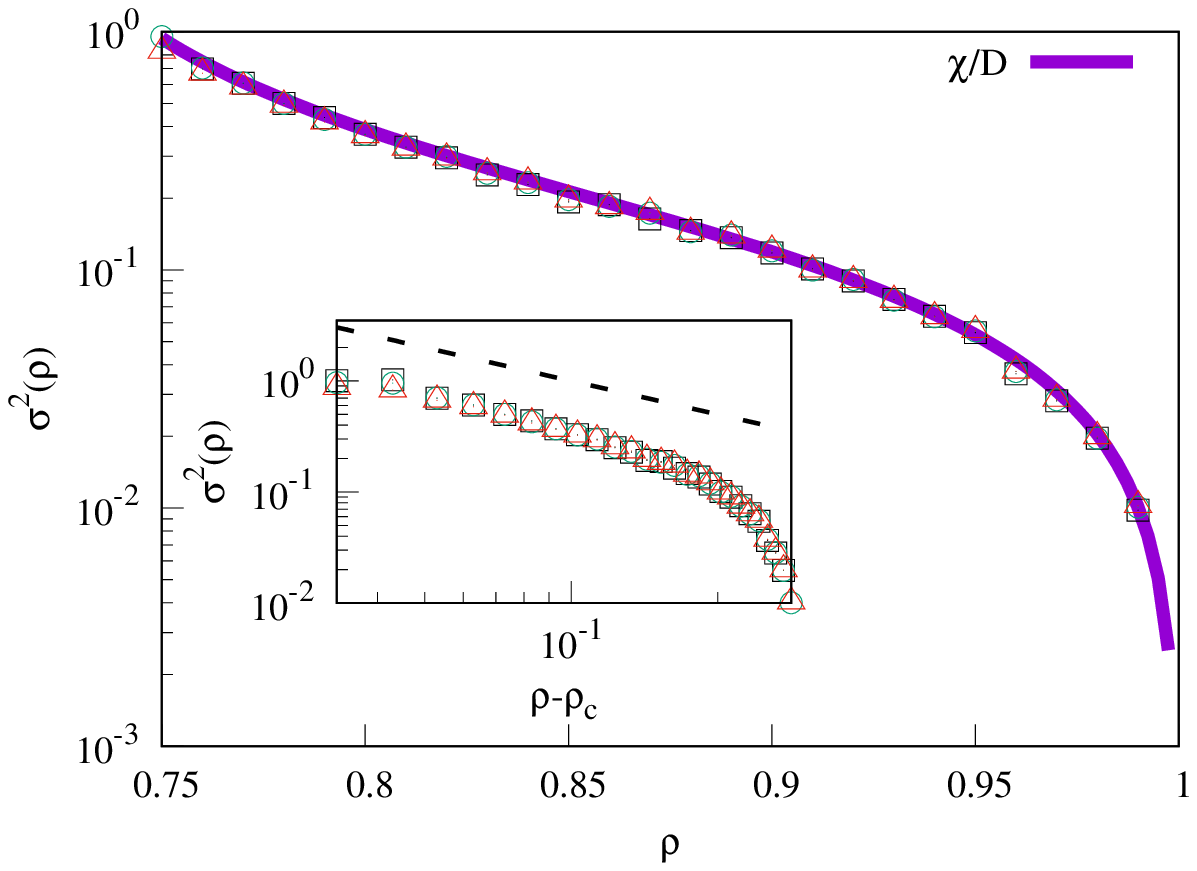}}
\caption{ {Verification of Einstein relation Eq. (\ref{ER1}) in gLHM.} Scaled variance $\sigma^2(\rho)$ of subsystem particle-number obtained from simulations (points) and the ratio $\chi(\rho)/D(\rho)$ of transport coefficients obtained from hydrodynamic theory (lines) is plotted as a function of density $\rho$. Top panel - finite range hopping ($l=2$) and bottom panel - infinite range hopping ($l \rightarrow \infty$). For $l=2$ (top panel), system size $L=5000$ and subsystem of size $l_{sub}=50$; for $l\rightarrow \infty$ (bottom panel), $L=10000$ and subsystem sizes $l_{sub}=50$ (red triangles), $100$ (green circles) and $200$ (black rectangles); we throughout take $p=q=1/2$ and thus $\rho_c =1/\sqrt{2}$. Inset, bottom panel: We plot scaled variance $\sigma^2(\rho)$ as a function of $(\rho-\rho_c)$, where the guiding dashed line shows the simple-pole singularity $\sigma^2 \sim (\rho-\rho_c)^{-1}$ as predicted by Einstein relation Eq. (\ref{ER1}).}
\label{ER_plot}
\end{figure}

Therefore, for a diffusive system on a periodic domain, which is the case for generalized long-hop model on a ring, macroscopic fluctuation theory predicts an equilibrium-like Einstein relation between the ratio of the transport coefficients and the number fluctuation. Next we verify Einstein relation for gLHM with finite as well as infinite range hopping.
To this end, in Fig. \ref{ER_plot}, we plot scaled variance $\sigma^2(\rho)$ of subsystem particle-number, obtained from simulations (points), and theoretically obtained ratio $\chi(\rho)/D(\rho)$ of the conductivity and the bulk-diffusion coefficient (lines) as a function of density $\rho$ for $p=q=1/2$ and for finite ($l=2$, top panel) and infinite ($l \rightarrow \infty$, bottom panel) range hopping. 
To check Einstein relation for $l=2$, we use the numerical values of the bulk-diffusion coefficient $D(\rho)$ as a function of density using Eq. (\ref{D l2}) and the conductivity
\begin{eqnarray}
\label{chi-2}
\chi(\rho) = \frac{q}{2} {\cal B}^{(3)}(\rho) + \frac{p}{2} {\cal B}^{(2)}(\rho) + 2q \left[ {\cal A}^{(2)}(\rho) - {\cal A}^{(3)}(\rho) \right], \nonumber 
\end{eqnarray}
using Eq. (\ref{chi-l})], where ${\cal B}^{(3)}(\rho) = \rho P(g=1|\rho)$, ${\cal B}^{(2)}(\rho) = \rho P(g=0|\rho)$, ${\cal A}^{(2)}(\rho) = P(g \ge 2|\rho)$ and ${\cal A}^{(3)}(\rho) = P(g \ge 3|\rho)$ are directly evaluated as a function of density from simulations by calculating the gap distribution $P(g|\rho)$. 
For $l \rightarrow \infty$, we use the analytic expressions in Eqs. (\ref{D-inf}) and (\ref{chi-inf}) to calculate the ratio of the transport coefficients $\chi(\rho)/D(\rho)$. 
In both cases of finite and infinite range hopping, we find that Einstein relation as in Eq. (\ref{ER1}) is quite well satisfied, thus establishing a direct connection between number fluctuation and transport in generalized long-hop model.   
Note that, though it remains finite for any density and activity in the case of finite range hopping, number fluctuation diverges in the case of infinite range hopping upon approaching critical density $\rho_c=1/\sqrt{2}$, and remains diverging beyond. In inset of bottom panel, Fig. \ref{ER_plot}, we plot the scaled number fluctuation $\sigma^2(\rho)$ for infinite range hopping as a function of $\Delta=(\rho-\rho_c)$, demonstrating that, similar to the conductivity, the number fluctuation also has a simple-pole singularity $\sigma^2 \sim \Delta^{-1}$, as predicted by Einstein relation.

\begin{figure}[tb!]
\centering
{\includegraphics[width=1.0 \linewidth]{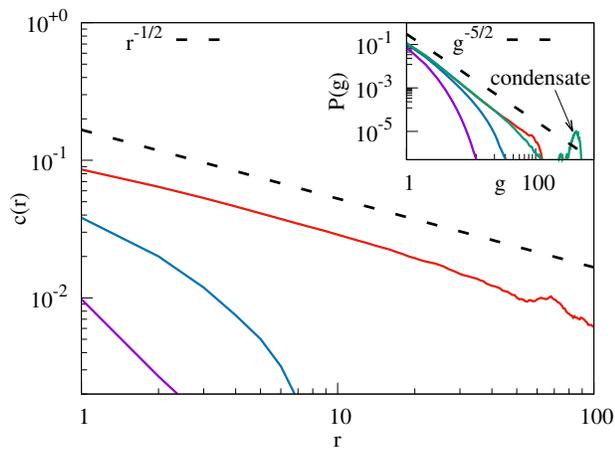}}
\caption{ Two-point correlation function $c(r) = (\langle \eta_i \eta_{i+r} \rangle - \rho^2)$ is plotted as a function of distance  $r$ is plotted for densities $\rho=0.9$ (magenta), $0.8$ (sky-blue), and $0.71$ (red, near criticality). The power-law tail of $c(r) \sim r^{-1/2}$ near critical point $\rho = 0.71$ demonstrates the presence of long-ranged spatial correlation in the system. Inset: Gap distribution $P(g)$ is plotted as a function of gap size $g$ is for densities $\rho=0.9$ (magenta), $0.8$ (sky-blue), $0.71$ (red, near criticality) and $0.5$ (green).   }
\label{cr_plot}
\end{figure}

\subsection{``Superfluid'' transition and ``giant'' number fluctuation for infinite range hopping}
\label{sec-gnf}

As shown in Sec. \ref{sec-IRH}, generalized long-hop model with infinite range hopping undergoes a ``superfluid'' transition beyond a critical value of density $\rho \le \rho_c=\tilde q^{1/2}$ or beyond a scaled activity $\tilde q \ge \tilde q_c = \rho^2$ where $\tilde q = q/(p+q)$. It is quite instructive to analyse the nature of the phase transition through the dynamics of gaps or vacancies in the light of a previously studied mass aggregation model \cite{Barma_PRL1998}, which is the unbounded version of gLHM with infinite range hopping. Let us consider two adjacent gaps, which are separated by a particle and subsequently coalesce into a larger gap as the particle performs a long range hop. The impossibility of the reverse process corresponding to the above event is precisely the reason for the violation of detailed balance in the system. Clearly the above mentioned irreversible nature of coalescence process facilitates formation of gaps of larger sizes. However, short range hops reduce a gap as particles on either side of a gap invade inside it and thus fragment a vacancy cluster. For $\rho < \rho_c = {\tilde q}^{1/2}$, the system organizes itself in such way that long-hops win over short-hops, leading to the formation of a macroscopic gap - a ``condensate'' of vacancies or holes. Due to the formation of a macroscopic size vacancy cluster, translational symmetry of the system breaks down and accordingly the phase is called an ``ordered'' one. On the other hand, for $\rho \ge \rho_c$ (or $q \le \tilde q_c$), the system remains homogeneous and the corresponding phase is called a ``disordered'' one.

Importantly, the simple-pole singularity in the conductivity $\chi(\rho)$ has the following implications.
\\
\\
(i) {\it Simple-pole singularity in number fluctuation.$-$} As shown in bottom panel of Fig. \ref{ER_plot} for infinite range hopping, particle-number fluctuation diverges near criticality (and remains diverging beyond). This can be understood from Einstein relation $\sigma^2(\rho) = \chi(\rho)/D(\rho)$ as in Eq. (\ref{ER1}), which immediately implies a simple pole in the scaled subsystem particle-number fluctuation $\sigma^2(\rho) \sim (\rho - \rho_c)^{-1}$, leading to diverging particle-number fluctuation near criticality (see inset of Fig. \ref{ER_plot}).
\\
\\
(ii) {\it Gap distribution.$-$} For density $\rho < \rho_c$ (or $\tilde q > \tilde q_c$), the system exhibits bimodal gap distribution in the ordered phase, signifying a condensation transition (CT) in the system \cite{Barma_PRL1998}. Indeed, following Ref. \cite{Das} and using fluctuation-response relation Eq. (\ref{FR}), we can show that the simple-pole singularity structure of scaled number fluctuation $\sigma^2(\rho) \sim (\rho - \rho_c)^{-1}$ implies a gap distribution $P(g) \sim {\rm const.} g^{-5/2} + {\rm const.} \delta[g-(\rho' - \rho_c') N]$ with $\rho'$ being corresponding mass density in UgLHM as in Eq. (\ref{rho-rhop}), having a power-law tail near criticality and a ``delta-peak'' at a macroscopic gap of size $(\rho' - \rho_c') N$ in the ordered phase \cite{Barma_PRL1998}; see the Appendix Sec. D for details. In inset of Fig. \ref{cr_plot}, we plot gap distribution $P(g)$ as a function of gap size $g$ for various densities $\rho=0.9$ (magenta), $0.8$ (sky-blue), $0.71$ (red, near criticality) and $0.5$ (green, ordered phase); one can see the power-law tail near criticality (red line) and beyond criticality ($\rho < 1/\sqrt{2}$), there appears, in addition to the power-law tail, a peak at a large gap, indicating the formation of a macroscopic size vacancy (hole) cluster in the system. 
Notably, the transition from the homogeneous fluid phase to the translational-symmetry-broken ordered phase, a coexisting phase of two distinct densities, is of first-order.
\\
\\
(iii) {\it Long-ranged spatial correlations.$-$} Not surprisingly, as the system becomes infinitely conducting, long-ranged spatial correlations is expected to build up in the system; in fact, they are present near the  critical point as well as in the ordered phase. For gLHM near criticality, following Refs. \cite{Das} and \cite{Jain}, we calculate the two-point density correlation $c(r) = (\langle \eta_i \eta_{i+r} \rangle - \rho^2) \sim r^{-1/2} \exp(-r/\xi)$, which has a power-law tail with exponent $1/2$ and a cut-off distance, called the correlation length, $\xi \sim (\rho-\rho_c)^{-\nu}$ where $\nu=2$ (shown below by using simple scaling arguments); also see Appendix Sec. E for calculation details. In Fig. \ref{cr_plot}, we have plotted correlation function $c(r)$ as a function of distance $r$ for various densities $\rho=0.9$ (magenta), $0.8$ (sky-blue), and $0.71$ (red). Theoretically obtained $r^{-1/2}$ power-law tail (red line) in the correlation function $c(r)$ agrees quite well with that obtained from simulations (points).
\\
\\
(iv) {\it Giant number fluctuations.$-$} Due to the presence of the long-ranged correlations, there are ``giant'' number fluctuations in the system near criticality as well in the ordered phase, where infinite conductivity, through Einstein relation, immediately implies diverging fluctuation [as discussed in point (i)]. Indeed, for infinite range hopping, we obtain below, by using simple scaling arguments, the near critical standard deviation of subsystem particle-number $\sqrt{\langle (\Delta n)^2 \rangle} \sim \langle n \rangle^{\alpha'}$ where $\alpha' > 1/2$. The diverging fluctuations persist in the bulk even in the ordered phase because changing the global density in the ordered phase affects only the condensate size whereas the density of the bulk ``superfluid'' remains the same. 
Indeed the power-law form of the density correlation can be related to the diverging number fluctuation as following. Near criticality as $\Delta  = (\rho-\rho_c) \rightarrow 0$ and correlation length $\xi \sim \Delta^{-\nu}$, the scaled variance $\sigma^2(\rho)$ is related to the integrated correlation function as $\sigma^2(\rho) \simeq \int c(r) dr \sim \int_0^{\xi} r^{-1/2} dr \sim \xi^{1/2} \sim \Delta^{-\nu/2}$. Now explicitly using simple-pole singularity of scaled number fluctuation $\sigma^2 \sim \Delta^{-1} \sim \Delta^{-\nu/2}$, we obtain $\nu = 2.$
Moreover, taking the correlation length $\xi \sim l_{sub} \gg 1$ (but, assuming $l_{sub} \ll L$), the standard deviation of subsystem particle-number $n$ in a subsystem of length $l_{sub}$ is given by $\sqrt{\langle (\Delta n)^2 \rangle} = \sqrt{(\langle n^2 \rangle - \langle n \rangle^2)} \sim \sqrt{l_{sub} \xi^{1/2}} \sim (l_{sub})^{\alpha'}$, leading to the scaling exponent $\alpha'=3/4$ corresponding to ``giant'' number fluctuation with the standard deviation $\sqrt{\langle (\Delta n)^2 \rangle} \sim \langle n \rangle^{\alpha'}$.
However, when the subsystem size $l_{sub} \sim L$ is of order system size, there is a strong finite-size effect and the correlation length, which is related to the width of mass fluctuation in UgLHM, has a scaling $\xi \sim L^{2/3}$ \cite{Evans}. Interestingly, in that case, the standard deviation of subsystem particle-number scales slightly differently and crosses over to the following scaling of the giant number fluctuation $\sqrt{\langle (\Delta n)^2 \rangle} \sim \sqrt{L \xi^{1/2}} \sim L^{\alpha'}$ with exponent $\alpha'=2/3$.  
As evident from the above discussions, diverging conductivity, through Einstein relation between the transport coefficients and fluctuation, is intimately connected to the giant number-fluctuation in the system.

\section{Summary and concluding remarks}
\label{sec-sum}

We have derived hydrodynamics of a prototypical model of self-propelled particles, called generalized long-hop model (gLHM). 
The model-system consists of hardcore particles on a lattice with a ring geometry. A particle hops symmetrically a variable distance, chosen from a probability distribution, provided there is an empty lane (a cluster of vacancies or holes) in front of it; if the chosen hop-length at any time is greater than the length of the empty lane, also called gap, the particle sits just at the end of the lane adjacent to the nearest particle in that hopping direction. 
The dynamics captures, albeit in a crude way, the ballistic motion in self-propelled particles, such as bacteria  \cite{Ariel_NatCom2015, Matthaus}, which, due to persistence,  traverse a relatively long distance in a straight stretch during a typical time interval $\tau_0$ - the persistence time of the individual bacteria. Clearly, on a time scale $\tau_0$, a self-propelled particle can be simply considered to have made a ``long range hop'' of distance $l= u \tau_0$, with $u$ being the typical speed of the particles. For generic parameter values, the model violates detailed balance, which is manifest in the formation of larger vacancy clusters. Indeed, an entire vacancy cluster can move in unison and merge with neighboring vacancies to create an even larger cluster, thus bringing in clustering, or cooperativity, in the system. The short-ranged hops on the other hand favor fragmentation of clusters and try to homogenize the system. In the limit of infinite-ranged hopping and beyond the critical values of density and activity, long-hops dominate over short-hop and vacancy clusters grow upto macroscopic size and form a ``condensate''.

To study the response of the system to a small external perturbation, we apply, following macroscopic fluctuation theory, a small force field, which couples to local particle-number by making forward (along the force field) and backward particle-hopping rates slightly asymmetric, i.e., the forward hopping rates being slightly more favorable than the backward ones. In that case, the large-scale hydrodynamic evolution of the local density is governed by two density-dependent transport coefficients - the bulk-diffusion coefficient $D(\rho)$  and the conductivity $\chi(\rho)$. When the typical long range hop-length is finite, the two transport coefficients remain finite as a function of density as well as activity. However, when  the typical long range hop-length diverges, as in the case of infinite range hopping considered in Sec. \ref{sec-IRH}, the system organizes itself in such a way that it undergoes a ``superfluid'' transition upon tuning global density and activity. Upon approaching the superfluid phase, say by tuning the activity, the conductivity starts diverging and the ``giant'' number fluctuations appear in the system; note that the diffusivity remains finite. Upon increasing the activity further into the ordered phase, diverging conductivity and giant-number number fluctuations both persist. Moreover, we demonstrate that, in both cases of finite and infinite range hopping, the scaled subsystem number-fluctuation $\sigma^2(\rho)$ is related to the ratio of the transport coefficients, through an Einstein relation $\sigma^2(\rho)=\chi(\rho)/D(\rho)$. Indeed, for infinite range hopping, the Einstein relation establishes a direct connection between the diverging conductivity and the ``giant'' number fluctuation in the system. 
Furthermore, we argue through Einstein relation how the singularity of the conductivity implies a condensation transition (CT) in the system.

It is quite remarkable that the above mentioned superfluid transition, somewhat contrary to the naive expectation, is actually induced by an instability in the conductivity, not by the usual diffusive instability. In that sense, the phase transition is truly a nonequilibrium one and, in the classical regime, presumably does not have any equilibrium counterpart. Thus the hydrodynamic theory of the generalized long-hop model provides a definitive, but hitherto unanticipated, unified mechanism of anomalous transport and fluctuations, both of which together, being typical of self-propelled particles, characterize long-ranged correlations in the system. In retrospect, in a self-propelled particle system \cite{Goldstein_PRL2004}, perhaps it is not difficult to visualize large mobile masses and the resulting long-ranged velocity correlations setting in the system. Indeed, the coherent motion of masses, as incorporated in the dynamical rules of gLHM, helps the system organizing into an infinitely conducting state, where the large mobility of masses contributes to the  enhancement of the ``compressibility'' [encoded in Eqs. (\ref{FR}) and (\ref{ER1})], resulting in ``giant'' number fluctuations in the system. The analytic results of this paper lends credence to the scenario that the ``giant'' number fluctuations in active-matter systems may actually be governed by an underlying superfluid structure.

We believe our study could be relevant in the context of collective behaviors of micro-organisms as in a bacterial colony. It is interesting to note that superfluidlike transitions have been observed in the past experimentally in bacterial suspensions \cite{Clement}, where viscous resistance to the bacteria moving in the surrounding solvent medium is shown to be highly reduced. Though the bacterial superfluidity is influenced by various fluid dynamical interactions between swimming bacteria and the solvent \cite{Hatwalne_PRL2004}, our hydrodynamic theory could provide another route to understanding the experimentally observed reduced viscosity in bacterial colony.

The generalized long-hop model elucidates not only the delicate interplay between persistence and interactions in self-propelled particles, but also connects to the paradigmatic mass-aggregation processes, which have been studied intensively in the past \cite{Krapivsky_PRE1996, Barma_PRL1998, MAM}.  
Furthermore, to the question of what the signature of condensation or superfluid transition would be in a self-propelled particle system, we propose the following: (a) Bimodal cluster distribution (of holes or particles) having a power-law tail and a peak at a macroscopic value and (b) the conductivity having a pole at a critical density and diverging beyond. These considerations indeed leave open the intriguing possibility of characterizing a broad class of self-propelled particles through the characteristics of ``superfluidlike'' transition \cite{Bar_PRE2006, Levis_PRE2014, Barberis, Golestanian_PRE2019}. Overall, we believe our analytic results  provide a fresh perspective to the collective behaviors in self-propelled particle systems, which are often seen in the light of motility-induced phase separation \cite{Cates_PRL2008}.

\section*{Acknowledgement} 

We thank Mustansir Barma, Sakuntala Chatterjee, Rahul Dandekar, Subhrangshu S. Manna and R. Rajesh for discussions. P.P. gratefully acknowledges the Science and Engineering Research Board (SERB), India, under Grant No. MTR/2019/000386, for financial support. T.C. acknowledges research fellowship [grant no. 09/575 (0124)/2019-EMR-I] from the Council of Scientific and Industrial Research (CSIR), India.

\section*{APPENDIX} 

Here we provide calculation details of deriving the loss and gain rates at a particular site for short and long range hopping.

\subsection{gLHM with hop-length distribution $\phi(l')=\delta_{l',l}$ having finite hopping range $l$}
\label{A-frh}

\subsubsection{Hydrodynamics}

In order to calculate conductivity in the hydrodynamic time-evolution equation of density field, we  bias the system by applying a small force of magnitude $F$, say, in anti-clockwise direction. Due to the presence of a small biasing force $F$, hopping rates are modified. We denote the modified long-hop rates as $q^F_R(l)$ and $q^F_L(l)$  and the modified short-hop rates as $p^F_R$ and $p^F_L$, where the subscripts ``R'' and ``L'' denote  anti-clockwise (in the direction of the biasing force) and clock-wise (opposite to biasing force) directions, respectively. We explicitly write the lattice spacing $a$ in the equations below.

\subsubsection{Gap size $g<l$}

When gap size $g<l$ where $l$ is the maximum possible hop-length, the mass loss and gain in this case (including the short hop) are shown below through all possible hopping events and the corresponding gain and loss rates. \\
The rate of loss of mass from site $i$ due to hopping of length $g$ and short hop to the right 
\begin{eqnarray}
J^{-}_{R,<} (i\rightarrow{i+g}) 
&=& \frac{q^F_R(g)}{2}  \langle \eta_{i}\overline{\eta}_{i+1} \dots \overline{\eta}_{i+g}\eta_{i+g+1} \rangle 
\nonumber \\
&& + \frac{p^F_R}{2}  \langle \eta_{i}\overline{\eta}_{i+1} \rangle
\nonumber \\
&\simeq & \frac{q^F_R(g)}{2} \mathcal{B}_{i+g+1}^{(g+2)} + \frac{p^F_R}{2} \langle \eta_{i}(1-\eta_{i+1}) \rangle, 
\nonumber 
\end{eqnarray}
where we denote the correlation $\langle \eta_{i-1}\overline{\eta}_{i} \overline{\eta}_{i+1} \dots \overline{\eta}_{i+g-1}\eta_{i+g} \rangle \equiv \mathcal{B}_{i+g}^{(g+2)}.$ For this particular hopping event to be possible, the following conditions to be satisfied: (i) Site $i$ must be occupied, (ii) the sites from $(i+1)$ upto $(i+g)$ must be vacant (so that the particle can jump a distance $g$) and (iii) then the site $(i+g+1)$ must be occupied [so that the particle does not jump beyond site $(i+g)$]. Similarly, other possible gain and loss terms can be constructed as given below.\\
The rate of gain of mass from site $i+g$ due to hopping of length $g$ and short hop to the left,
\begin{eqnarray}
J^{+}_{L,<} (i+g \rightarrow i)
&=& \frac{q^F_L(g)}{2} \langle \eta_{i-1} \overline{\eta}_{i} \dots \overline{\eta}_{i+g-1}\eta_{i+g} \rangle
\nonumber \\
&& + \frac{p^F_L }{2} \langle \overline{\eta}_{i}\eta_{i+1} \rangle
\nonumber \\
&\simeq & \frac{1}{2} q^F_L(g) \mathcal{B}_{i+g}^{(g+2)} + \frac{1}{2} p^F_L  \langle (1-\eta_i)\eta_{i+1} \rangle
\nonumber 
\end{eqnarray}
The rate of loss of mass from site $i$ due to hopping of length $g$ plus short hop to the left,
\begin{eqnarray}
J^-_{L,<}(i\rightarrow{i-g})
&=& \frac{q^F_L(g)}{2}  \langle \eta_{i-g-1}\overline{\eta}_{i-g} \dots \overline{\eta}_{i-1} \eta_{i} \rangle 
\nonumber \\
&& + \frac{p^F_L }{2} \langle \overline{\eta}_{i-1}\eta_{i} \rangle
\nonumber \\
&=& \frac{1}{2} q^F_L(g) \mathcal{B}_{i}^{(g+2)} + \frac{1}{2} p^F_L \langle (1-\eta_{i-1})\eta_{i} \rangle
\nonumber
\end{eqnarray}
The rate of gain of mass at site $i$ due to hopping of length $g$ and short hop to the right,
\begin{eqnarray}
J^+_{R,<}(i-g \rightarrow i)
&=& \frac{q^F_R(g)}{2}  \langle \eta_{i-g}\overline{\eta}_{i-g+1} \dots \overline{\eta}_{i} \eta_{i+1} \rangle 
\nonumber \\
&& + \frac{p^F_R }{2} \langle \eta_{i-1}\overline{\eta}_{i} \rangle
\nonumber \\
&=& \frac{1}{2} q^F_R(g))\mathcal{B}_{i+1}^{(g+2)}  + \frac{1}{2} p^F_R  \langle \eta_{i-1}(1-\eta_i) \rangle
\nonumber
\end{eqnarray}  
Therefore, net rate of change of mass at site $i$ due to the above hopping processes can be written as
\begin{widetext}
\begin{eqnarray}
J^{+}_{L,<}(l) - J^{-}_{R,<}(l) + J^+_{R,<}(l) - J^-_{L,<}(l) = \sum_{g=1}^{l-1} \left[  J^{+}_{L,<} (i+g \rightarrow i) - J^{-}_{R,<} (i \rightarrow {i+g}) + J^+_{R,<}(i-g \rightarrow i) - J^-_{L,<}(i\rightarrow{i-g}) \right] 
\nonumber \\
= - \frac{q}{2} \sum_{g=1}^{l-1} \left[ \left\lbrace (1 + \frac{Fga}{2}) \mathcal{B}_{i+g+1}^{(g+2)}-(1-\frac{Fga}{2}) \mathcal{B}_{i+g}^{(g+2)}\right\rbrace - \left\lbrace (1+\frac{Fga}{2}) \mathcal{B}_{i+1}^{(g+2)}-(1-\frac{Fga}{2}) \mathcal{B}_{i}^{(g+2)} \right\rbrace \right]
\nonumber \\
- \frac{p}{2} \left[ \left(1+\frac{Fa}{2} \right) \langle \eta_{i}(1-\eta_{i+1}) \rangle + \left(1 - \frac{Fa}{2} \right) \langle (1-\eta_{i-1})\eta_{i} \rangle - \left(1 - \frac{Fa}{2} \right) \langle (1-\eta_i)\eta_{i+1} \rangle - \left(1 + \frac{Fa}{2} \right) \langle \eta_{i-1}  (1-\eta_i) \rangle  \right]
\nonumber \\
= - \frac{q}{2} \sum_{g=1}^{l-1} \left[ \dots \right]
+ \frac{p}{2} [\langle \eta_{i+1} \rangle - 2 \langle \eta_i \rangle + \langle \eta_{i-1} \rangle] + \frac{p}{2} \frac{Fa}{2} \left[ - (\langle \eta_{i+1} \rangle -  \langle \eta_{i-1}) \rangle) + 2 (\langle \eta_i \eta_{i+1} \rangle - \langle \eta_{i-1} \eta_i \rangle) \right] \nonumber \\
\end{eqnarray}
\end{widetext}
By denoting $\langle \eta_i \rangle = \rho_i$ and $\langle \eta_{i-1} \eta_i \rangle = \mathcal{B}_{i}^{(2)}$, we obtain the net flux corresponding to the events when $g<l$ as given below,
\begin{widetext}
\begin{eqnarray}
J^{+}_{L,<}(l) - J^{-}_{R,<}(l) + J^+_{R,<}(l) - J^-_{L,<}(l) = \frac{p}{2} [\rho_{i+1} - 2 \rho_i + \rho_{i-1}] - \frac{p}{2} \frac{Fa}{2} \left[ (\rho_{i+1} -  \rho_{i-1}) - 2 \left\{  \mathcal{B}_{i+1}^{(2)} -  \mathcal{B}_{i}^{(2)} \right\} \right] 
\nonumber \\
 - \frac{q}{2} \sum_{g=1}^{l-1} \left[ \left\lbrace (1 + \frac{Fga}{2}) \mathcal{B}_{i+g+1}^{(g+2)}-(1-\frac{Fga}{2}) \mathcal{B}_{i+g}^{(g+2)}\right\rbrace - \left\lbrace (1+\frac{Fga}{2}) \mathcal{B}_{i+1}^{(g+2)} -(1-\frac{Fga}{2}) \mathcal{B}_{i}^{(g+2)} \right\rbrace \right].
\end{eqnarray}
\end{widetext}
We now perform a small-gradient ${\cal O}(1/L)$ Taylor series expansion as given below,
\begin{eqnarray}
 \mathcal{B}_{i+1}^{(k)} &\equiv & \nonumber \mathcal{B}^{(k)}(x+1/L) \\ &\simeq & \mathcal{B}^{(k)}(\rho(x))  + \frac{1}{L} \partial_x \mathcal{B}^{(k)}(\rho(x))  + \frac{1}{2 L^2} \partial^{2}_x \mathcal{B}^{(k)}(\rho(x)),
 \nonumber
\\
 \mathcal{B}_{i+g+1}^{(k)} &\equiv & \nonumber \mathcal{B}^{(k)}(x+(g+1)/L) \\ &\simeq & \mathcal{B}^{(k)}(\rho(x))  + \frac{(g+1)}{L} \partial_x \mathcal{B}^{(k)}(\rho(x)) 
 \nonumber \\
 && + \frac{(g+1)^{2}}{2 L^2} \partial^{2}_x \mathcal{B}^{(k),}(\rho(x)) ,
 \nonumber
\\
 \mathcal{B}_{i+g}^{(k)} &\equiv & \nonumber \mathcal{B}^{(k)}(x+g/L) \\ &\simeq & \mathcal{B}^{(k)}(\rho(x))  + \frac{g}{L} \partial_x \mathcal{B}^{(k)}(\rho(x))  + \frac{g^{2}}{2 L^2} \partial^{2}_x \mathcal{B}^{(k)}(\rho(x)).
 \nonumber
\end{eqnarray}

In the diffusive scaling limit $i \rightarrow x=i/L$, $t \rightarrow t/L^2$ and $a \rightarrow 1/L$, we can recast the net rate of change of mass in the form of the divergence of a current upto ${\cal O}\left( 1/L^2 \right)$,
\begin{equation}
J^{+}_{L,<} - J^{-}_{R,<} + J^+_{R,<} - J^-_{L,<} = \frac{\partial}{\partial x} \left[ D_{<}(\rho) \frac{\partial \rho}{\partial x} - \chi_{<}(\rho) F\right] 
\end{equation}
where we denote
\begin{eqnarray}
D_{<}(\rho)&=& - \frac{q}{2} \frac{\partial}{\partial \rho}\left[ \sum_{g=1}^{l-1} g \mathcal{B}^{(g+2)}(\rho) \right] +\frac{p}{2} \nonumber \\ & = &- \frac{q}{2} \frac{\partial}{\partial \rho}\left[ \rho \sum_{g=1}^{l-1} g P(g|\rho') \right] +\frac{p}{2},
\\
\chi_{<}(\rho)&=&\frac{q}{2} \sum_{g=1}^{l-1} g^{2} \mathcal{B}^{(g+2)}(\rho)  + \frac{p}{2}\mathcal{B}^{(2)}(\rho) \nonumber \\ & = & \frac{q}{2} \rho \sum_{g=1}^{l-1} g^{2} P(g|\rho')  + \frac{p}{2} \mathcal{B}^{(2)}(\rho),
\end{eqnarray}
by noting that the correlation $\mathcal{B}^{(g+2)}(\rho)$ in gLHM with density $\rho$ is related to the single-site mass distribution $P(g|\rho')$ in UgLHM with density $\rho'=1/\rho-1$.

\subsubsection{Gap size $g \geq l$ }

 Let us first consider the case for which the gap in front of $i$ th site is larger than or equal to $l$. In that case maximum hop length is $l$. If a particle hops from $i$th site to $(i+l)$th site, the corresponding mass-loss rate is given by
\begin{eqnarray}
  J^{-}_{R,\ge}(i \rightarrow i+l) & = & \frac{q}{2}(1+\frac{Fla}{2})\langle \eta_{i} \overline{\eta}_{i+1} \overline{\eta}_{i+2} \dots \overline{\eta}_{i+l}\rangle \nonumber \\ & = & \frac{q}{2}(1+\frac{Fla}{2})\langle (1-\overline{\eta}_{i}) \overline{\eta}_{i+1} \overline{\eta}_{i+2} \dots \overline{\eta}_{i+l}\rangle, \nonumber \\
 \end{eqnarray}
where we denote $\overline{\eta}_i = (1-\eta_i)$.

Similarly if a particle hops from $(i+l)$th to $i$th site site, the corresponding mass-gain rate is given by
\begin{eqnarray}
     J^{+}_{L,\ge}(i+l \rightarrow i)& = &\frac{q}{2}(1-\frac{Fla}{2})\langle \eta_{i+l} \overline{\eta}_{i+l-1}\overline{\eta}_{i+l-2} \dots \overline{\eta}_{i}\rangle \nonumber \\ & = & \frac{q}{2}(1-\frac{Fla}{2})\langle (1-\overline{\eta}_{i+l}) \overline{\eta}_{i+l-1}\overline{\eta}_{i+l-2} \dots \overline{\eta}_{i} \rangle. \nonumber \\
 \end{eqnarray}
 The net gain rate from the right side of $i$th site is written as, by adding the above two terms,
 \begin{widetext}
 \begin{eqnarray}
J^{+}_{L,\ge}(i+l \rightarrow i) - J^{-}_{R,\ge}(i \rightarrow i+l) &=& \nonumber \frac{q}{2}\{\langle \overline{\eta}_{i+l-1}\overline{\eta}_{i+l-2} \dots \overline{\eta}_{i}\rangle  - \langle \overline{\eta}_{i+1} \overline{\eta}_{i+2} \dots \overline{\eta}_{i+l}\rangle \} 
-\frac{qFla}{4} \{\langle \overline{\eta}_{i+l-1}\overline{\eta}_{i+l-2} \dots \overline{\eta}_{i}\rangle  
\nonumber \\
&& + \langle \overline{\eta}_{i+1} \overline{\eta}_{i+2} \dots \overline{\eta}_{i+l}\rangle 
- 2 \langle \overline{\eta}_{i} \overline{\eta}_{i+1} \overline{\eta}_{i+2} \dots \overline{\eta}_{i+l}\rangle \}
\nonumber \\
&=&  \frac{q}{2}[\mathcal{A}_{i+l-1}^{(l)} - \mathcal{A}_{i+l}^{(l)}] - \frac{qFla}{4}[\mathcal{A}_{i+l-1}^{(l)} + \mathcal{A}_{i+l}^{(l)} - 2 \mathcal{A}_{i+l}^{(l+1)}],
 \end{eqnarray}
 \end{widetext}
Similarly, the net gain rate from the left side of site $i$ can be written a
 \begin{eqnarray}
 J^{+}_{R,\ge}(i-l \rightarrow i) - J^{-}_{L,\ge}(i \rightarrow i-l) = \frac{q}{2}[\mathcal{A}_{i}^{(l)} -\mathcal{A}_{i-1}^{(l)}] \nonumber \\ +\frac{qFla}{4}[\mathcal{A}_{i}^{(l)}+\mathcal{A}_{i-1}^{(l)} - 2\mathcal{A}_{i}^{(l+1)}].
 \end{eqnarray}
Therefore the net rate of change of mass at site $i$ due to the above processes can be written by adding the above four terms, which, in the diffusive scaling limit $i \rightarrow x = {i}/{L}$, $t \rightarrow {t}/{L^{2}}$ and lattice spacing $a \rightarrow 1/L$, reduces to the divergence of a hydrodynamic current upto $\mathcal{O}\left(\frac{1}{L^{2}}\right)$ ,
\begin{widetext}
\begin{eqnarray}
\nonumber J^{+}_{L,\ge} - J^{-}_{R,\ge} + J^{+}_{R,\ge} - J^{-}_{L,\ge}   &=& \frac{q}{2}[\{\mathcal{A}^{(l)}\left(x+(l-1)/L\right)-\mathcal{A}^{(l)}(x+l/L)\}-\{\mathcal{A}^{(l)}(x-1/L)-\mathcal{A}^{(l)}(x)\}] \\ \nonumber
   -\frac{qFla}{4}[\{\mathcal{A}^{(l)}(x+(l-1)/L)+\mathcal{A}^{(l)}(x+l/L)\}  & - & \{\mathcal{A}^{(l)}(x-1/L)+\mathcal{A}^{(l)}(x)\}
   - 2\{\mathcal{A}^{(l+1)}(x+l/L)-\mathcal{A}^{(l+1)}(x)\}]
\nonumber \\ 
   \Rightarrow J^{+}_{L,\ge} - J^{-}_{R,\ge} + J^{+}_{R,\ge} - J^{-}_{L,\ge}  &=& \frac{\partial}{\partial x} \left[ D_{\ge}(\rho)  \frac{\partial \rho}{\partial x} - \chi_{\ge}(\rho) F, \right]
 \label{events1}
\end{eqnarray}
   \end{widetext}
where we denote
\begin{eqnarray}
  D_{\ge}(\rho)=-\frac{ql}{2}\frac{\partial \mathcal{A}^{(l)}(\rho)}{\partial \rho}, \nonumber \\
  \chi_{\ge}(\rho)=\frac{q l^{2}}{2}\left[\mathcal{A}^{(l)}(\rho)-\mathcal{A}^{(l+1)}(\rho)\right],
  \nonumber
\end{eqnarray}
   and use Taylor series expansion,
   \begin{eqnarray}
   \mathcal{A}_{i+l}^{(l)} & \equiv & \mathcal{A}^{(l)}(x+l/L) \nonumber \\ & \simeq & \mathcal{A}^{(l)}(\rho(x)) + \frac{l}{L} \frac{\partial \mathcal{A}^{(l)}(\rho(x))}{\partial x} 
   \nonumber \\
   && + \frac{l^2}{2L^2} \frac{\partial^{2} \mathcal{A}^{(l)}(\rho(x))}{\partial x^{2}}  + \dots \nonumber,
   \end{eqnarray}
and
   \begin{eqnarray}
   \mathcal{A}_{i+l}^{(l+1)} & \equiv & \mathcal{A}^{(l+1)}(x+l/L) \nonumber \\ 
   & \simeq & \mathcal{A}^{(l+1)}(\rho(x)) + \frac{l}{L} \frac{\partial \mathcal{A}^{(l+1)}(\rho(x))}{\partial x} 
   \nonumber \\
  && + \frac{l^2}{2L^2} \frac{\partial^{2} \mathcal{A}^{(l+1)}(\rho(x))}{\partial x^{2}}  + \dots,
  \nonumber
   \end{eqnarray}
and similarly for the other terms.

\subsubsection{Continuity equation for local density}

Considering all possible hopping events and summing over the corresponding gap sizes, we obtain the hydrodynamic time-evolution for the density field $\rho(x,t)$,
\begin{widetext}
\begin{eqnarray}
\frac{\partial \rho(x,t)}{\partial t} = (J^{+}_{L,<} - J^{-}_{R,<} + J^{+}_{R,<} - J^{-}_{L,<}) + (J^{+}_{L,\ge} - J^{-}_{R,\ge} + J^{+}_{R,\ge} - J^{-}_{L,\ge}) = J^{+}_R + J^{+}_L - J^{-}_R - J^{-}_L 
\nonumber \\
=  \frac{q}{2} \left[ \{ {\cal A}^{(l)}_{i+l-1} - {\cal A}^{(l)}_{i+l} \} - \{ {\cal A}^{(l)}_{i-1} - {\cal A}^{(l)}_{i} \} \right] 
- \frac{qFl}{4} \left[ \{ {\cal A}^{(l)}_{i+l-1} + {\cal A}^{(l)}_{i+l} \} 
- \{ {\cal A}^{(l)}_{i-1} + {\cal A}^{(l)}_{i} \} - 2 \{ {\cal A}^{(l+1)}_{i+l} - {\cal A}^{(l+1)}_{i} \} \right]
 \nonumber \\ 
+ \sum_{l'=1}^{l-1} \frac{q}{2} \left[ \{ {\cal B}^{(l'+2)}_{i+1} - {\cal B}^{(l'+2)}_i \} - \{ {\cal B}^{(l'+2)}_{i+l'+1} - {\cal B}^{(l'+2)}_{i+l'} \} \right] 
- \sum_{l'=1}^{l-1} \frac{qFl'}{4} \left[ \{ {\cal B}^{(l+2)}_{i+l'+1} + {\cal B}^{l'+2}_{i+l'} \} - \{ {\cal B}^{(l'+2)}_{i+1} + {\cal B}^{(l'+2)}_i \} \right]
\nonumber \\
+ \frac{p}{2} \left[ \rho_{i+1} - 2 \rho_i + \rho_{i-1} \right] + \frac{pF}{4} \left[ \{ \rho_{i-1} - \rho_{i+1} \} + 2 \{ {\cal B}^{(2)}_{i+1} - {\cal B}^{(2)}_i \} \right]
\equiv - \frac{\partial}{\partial x} \left[ - D_{l}(\rho) \frac{\partial \rho}{\partial x} + \chi_{l}(\rho) F\right],
\end{eqnarray}
\end{widetext}
where we denote $J^+_R=J^+_{R,<}+J^+_{R,\ge}$, $J^{+}_L=J^{+}_{L,<}+J^{+}_{L,\ge}$, $J^{-}_R=J^{-}_{R,<}+J^{-}_{R,\ge}$ and $J^{-}_L=J^{-}_{L,<}+J^{-}_{L,\ge}$ and  the transport coefficients - the bulk-diffusion coefficient $D_l(\rho)$ and the conductivity $\chi_l(\rho)$, are given by
\begin{eqnarray}
   D_l(\rho)& = & D_{<}(\rho) + D_{\ge}(\rho) \nonumber \\ & = & \frac{p}{2}- \frac{q}{2} \frac{\partial}{\partial \rho}\left[ \sum_{g=1}^{l-1} g \mathcal{B}^{(g+2)}(\rho) \right]-\frac{ql}{2}\frac{\partial \mathcal{A}^{(l)}(\rho)}{\partial \rho}, \nonumber  \\
   \chi_l(\rho) &=& \chi_{<}(\rho) + \chi_{\ge}(\rho) \nonumber \\ &=& \frac{q l^{2}}{2}\left[\mathcal{A}^{(l)}(\rho)-\mathcal{A}^{(l+1)}(\rho)\right] +\frac{q}{2} \sum_{g=1}^{l-1} g^{2} \mathcal{B}^{(g+2)}(\rho)  
   \nonumber \\
   && + \frac{p}{2}\mathcal{B}^{(2)}(\rho),\nonumber 
\end{eqnarray}
constituting the first main results of the main text.

\subsection{Infinite-ranged gLHM with hop-length distribution $\phi(l')=\delta_{l',l}$ where $l \rightarrow \infty$}

In gLHM (exclusion version), a particle at a site $i$ hops to right or left (each direction chosen randomly with probability $1/2$) with slightly biased rates $q^F_R/2$ ($p^F_R/2$, depending on long or short hop) or $q^F_L/2$ ($p^F_L/2$), respectively, in the presence of a small biasing force $F$  according to the following rules. 
\\
\\
(A) Long hop:  With rate $q^F_R$ ($q^F_L$), a particle hops, without crossing any particle, symmetrically to right or left (with equal probability $1/2$) to the site adjacent to its {\it nearest occupied} site, i.e., it hops $g$ lattice spacing, provided its neighboring {\it empty} stretch has length $g$.
\\
\\
(B) Short hop: With rate $p^F_R$ ($p^F_L$), a particle hops to its right (left) nearest neighbor, provided the destination site is unoccupied. 
\\
\\
Below we consider all possible loss and gain terms and the corresponding rates with which a particle leaves or enters a site $i$.
The rate of loss of mass or particle from site $i$ due to hopping of length $g$ and short hop to the right,
\begin{eqnarray}
J^{-}_R (i\rightarrow{i+g})
&=& \frac{1}{2} q^F_R(l) \langle \eta_{i}\overline{\eta}_{i+1} \dots \overline{\eta}_{i+g}\eta_{i+g+1} \rangle 
\nonumber \\
&& + \frac{1}{2} p^F_R \langle \eta_{i}\overline{\eta}_{i+1} \rangle
\nonumber \\
&\simeq & \frac{1}{2} q^F_R(l) \mathcal{B}_{i+g+1}^{(g+2)} + \frac{1}{2} p^F_R \langle \eta_{i}(1-\eta_{i+1}) \rangle,  
\nonumber
\end{eqnarray}
The rate of gain of mass from site $i+g$ due to hopping of length $g$ and short hop to the left,
\begin{eqnarray}
J^{+}_L (i+g \rightarrow i)
&=& \frac{1}{2} q^F_L(g) \langle \eta_{i-1} \overline{\eta}_{i} \dots \overline{\eta}_{i+g-1}\eta_{i+g} \rangle 
\nonumber \\
&& + \frac{1}{2} p^F_L \langle \overline{\eta}_{i}\eta_{i+1} \rangle,
\nonumber \\
&\simeq & \frac{1}{2} q^F_L(g) \mathcal{B}_{i+g}^{(g+2)} + \frac{1}{2} p^F_L \langle (1-\eta_i)\eta_{i+1} \rangle.
\nonumber
\end{eqnarray}
The rate of loss of mass from site $i$ due to hopping of length $g$ plus short hop to the left,
\begin{eqnarray}
J^-_L(i\rightarrow{i-g})
&=& \frac{1}{2} q^F_L(g) \langle \eta_{i-g-1}\overline{\eta}_{i-g} \dots \overline{\eta}_{i-1} \eta_{i} \rangle 
\nonumber \\
&& + \frac{1}{2} p^F_L \langle \overline{\eta}_{i-1}\eta_{i} \rangle
\nonumber \\
&=& \frac{1}{2} q^F_L(g) \mathcal{B}_{i}^{(g+2)} + \frac{1}{2} p^F_L  \langle (1-\eta_{i-1})\eta_{i} \rangle
\nonumber
\end{eqnarray}
The rate of gain of mass at site $i$ due to hopping of length $g$ and short hop to the right,
\begin{eqnarray}
J^+_R(i-g \rightarrow i)
&=& \frac{1}{2} q^F_R(g) \langle \eta_{i-g}\overline{\eta}_{i-g+1} \dots  \overline{\eta}_{i} \eta_{i+1} \rangle 
\nonumber \\
&& + \frac{1}{2} p^F_R \langle \eta_{i-1}\overline{\eta}_{i} \rangle
\nonumber \\
&=& \frac{1}{2} q^F_R(g)\mathcal{B}_{i+1}^{(g+2)}  + \frac{1}{2} p^F_R \langle \eta_{i-1}(1-\eta_i) \rangle
\nonumber
\end{eqnarray}    
Therefore, net rate of change of mass at site $i$ can be written as
\begin{widetext}
\begin{eqnarray}
\frac{\partial \rho_i}{\partial t} = \sum_{g=1}^{\infty} \left[  J^{+}_L (i+g \rightarrow i) - J^{-}_R (i \rightarrow {i+g}) + J^+_R(i-g \rightarrow i) - J^-_L(i\rightarrow{i-g}) \right] 
\nonumber \\
= - \frac{q}{2} \sum_{g=1}^{\infty} \left[ \left\lbrace (1 + \frac{Fga}{2}) \mathcal{B}_{i+g+1}^{(g+2)}-(1-\frac{Fga}{2}) \mathcal{B}_{i+g}^{(g+2)}\right\rbrace - \left\lbrace (1+\frac{Fga}{2}) \mathcal{B}_{i+1}^{(g+2)}-(1-\frac{Fga}{2}) \mathcal{B}_{i}^{(g+2)} \right\rbrace \right]
\nonumber \\
- \frac{p}{2} \left[ \left(1+\frac{Fa}{2} \right) \langle \eta_{i}(1-\eta_{i+1}) \rangle + \left(1 - \frac{Fa}{2} \right) \langle (1-\eta_{i-1})\eta_{i} \rangle - \left(1 - \frac{Fa}{2} \right) \langle (1-\eta_i)\eta_{i+1} \rangle - \left(1 + \frac{Fa}{2} \right) \langle \eta_{i-1}  (1-\eta_i) \rangle  \right]
\nonumber \\
=- \frac{q}{2} \sum_{g=1}^{\infty} \left[ \left\lbrace (1 + \frac{Fga}{2}) \mathcal{B}_{i+g+1}^{(g+2)}-(1-\frac{Fga}{2}) \mathcal{B}_{i+g}^{(g+2)}\right\rbrace - \left\lbrace (1+\frac{Fga}{2}) \mathcal{B}_{i+1}^{(g+2)}-(1-\frac{Fga}{2}) \mathcal{B}_{i}^{(g+2)} \right\rbrace \right]
\nonumber \\
+ \frac{p}{2} [\rho_{i+1} - 2 \rho_i + \rho_{i-1}] - \frac{p}{2} \frac{Fa}{2} \left[ (\rho_{i+1} -  \rho_{i-1}) - 2 \left\{ \rho_{i+1} (1-{\cal c}(\rho_{i+1})) - \rho_i (1-{\cal c}(\rho_i)) \right\} \right],
\end{eqnarray}
\end{widetext}
where we simply denote $c(\rho)$ as the occupation probability in UgLHM corresponding to the density $\rho$ in gLHM.

We now perform a small-gradient ${\cal O}(1/L)$ Taylor series expansion as done in the previous sections (also see main text) and in the diffusive scaling limit $i \rightarrow x=i/L$ and $t \rightarrow t/L^2$, we finally obtain the hydrodynamic time-evolution of the density field $\rho(x,t)$,
\begin{equation}
\frac{\partial \rho(x,t)}{\partial t} = \frac{\partial}{\partial x} \left[ D(\rho) \frac{\partial \rho}{\partial x} \right] - \frac{\partial[\chi(\rho) F]}{\partial x}
\end{equation}
where the transport coefficients, the bulk-diffusion coefficient and the conductivity, are given by
\begin{eqnarray}
D(\rho)&=& - \frac{q}{2} \frac{d}{d \rho}\left[ \sum_{g=1}^{\infty} g \mathcal{B}^{(g+2)} \right] +\frac{p}{2}  \nonumber \\ & = &- \frac{q}{2} \frac{d}{d \rho}\left[ \rho \sum_{g=1}^{\infty} g P(g|\rho') \right] + \frac{p}{2} \nonumber \\ &=& - \frac{q}{2} \frac{d}{d \rho} \left( \rho \rho'(\rho) \right) + \frac{p}{2} = \frac{p+q}{2},
\label{D}
\\
\chi(\rho) &=& \frac{q}{2} \sum_{g=1}^{\infty} g^2 \mathcal{B}^{(g+2)}  + \frac{p}{2} \rho {\cal c}(\rho'(\rho)) \nonumber \\ &=& \frac{q}{2} \rho \sum_{g=1}^{\infty} g^2 P(g|\rho')  + \frac{p}{2} \rho {\cal c}(\rho'(\rho)) \nonumber \\ &=& \frac{q}{2} \rho \theta_2 (\rho'(\rho)) + \frac{p}{2} \rho {\cal c}(\rho'(\rho)).
\label{chi}
\end{eqnarray}
Here we have used the relation ${\cal B}^{(2)}(\rho) = \rho P(g=0|\rho') = \rho [1-{\cal c}(\rho'(\rho))]$ where 
$${\cal c}(\rho'(\rho))=\frac{\rho'(1-\rho'}{(1+\rho')}$$ 
is the probability that a site in UgLHM is occupied at a given density $$\rho'(\rho)= \sum_g g P(g|\rho') = \frac{1}{\rho}-1,$$ $\rho$ being the corresponding density in gLHM. 
For the derivation of occupation probability ${\cal c}(\rho')$ and second moment of mass $\theta_{2}(\rho'(\rho))$ as a function of density $\rho'$ in UgLHM, see the next section.
Finally, using relation between $c$, $\theta_{2}$ and $\rho'$ as a function of $\rho$ in Eqs. \ref{D} and \ref{chi}, we obtain the expressions for the two transport coefficients as function of density $\rho$,
\begin{eqnarray}
D(\rho)=\frac{p+q}{2} ~;~ \chi(\rho) = \frac{\rho(1-\rho)[(p+q) \rho^2 - 2q \rho + q]}{2[\rho^2-q/(p+q)]},
\nonumber
\end{eqnarray}
which constitute the second main results of the paper, reported in the main text.

\subsection{Calculation of second moment $\langle g_i^2 \rangle$ of local mass $g_i$ in UgLHM with $\phi(l')=\delta_{l'l}$ and $l \rightarrow \infty$}

We consider UgLHM with $\phi(l')=\delta_{l'l}$ and $l \rightarrow \infty$. We denote a configuration of UgLHM as a set of mass variables $\{g_i\}$. The dynamics in this special case of UgLHM consists of two processes chipping of a single unit of mass and diffusion of entire stack of mass from any site $i$. 
\\ \\
\textbf{Chipping}: With rate $p$, single unit mass at site $i$ is chipped off and transferred symmetrically to one of its nearest neighbor site with equal probability $1/2$:
$$
g_{i} \rightarrow g_{i}-1~;~ g_{i \pm 1} \rightarrow g_{i \pm 1}+1.
$$ 
\textbf{Diffusion}: With rate $q$, an entire stack of mass at site $i$ diffuse symmetrically to one of its nearest neighbor site with equal probability $1/2$:
$$
g_{i} \rightarrow 0~;~ g_{i \pm 1} \rightarrow g_{i \pm 1} + g_{i}.
$$ 
The occupancy of $i$th site is given by an indicator variable ${\cal c}_i$ at site $i$,
 \begin{equation}
 {\cal c}_{i}=(1-\delta_{g_{i},0}). \nonumber
 \end{equation}
The local mass variable $m_i(t)$ at site $i$ and at time $t$ evolves in an infinitesimal time-interval $dt$ according to the following stochastic dynamics:
\begin{eqnarray}
g_i(t+dt) =
\left\{
\begin{array}{ll}
g_i(t) + m_{i-1}(t)            & {\rm prob.}~ {\cal c}_{i-1} \frac{q}{2} dt, \\
g_i(t) + m_{i+1}(t)            & {\rm prob.}~ {\cal c}_{i+1} \frac{q}{2} dt, \\
g_i(t) - 1            & {\rm prob.}~ {\cal c}_{i}p  dt , \\
g_i(t) +1			  & {\rm prob.}~ {\cal c}_{i+1}\frac{p}{2}  dt , \\
g_i(t) +1			  & {\rm prob.}~ {\cal c}_{i-1}\frac{p}{2}  dt , \\
 0					  & {\rm prob.}~ {\cal c}_{i} q  dt , \\
g_i(t) 			 & {\rm prob.}~ 1-\Sigma dt, \\
\end{array}			 
\right.
\label{Longhop-unbiased}
\end{eqnarray}
with 
\begin{equation}
\Sigma = \left[ \frac{q}{2} {\cal c}_{i-1} + \frac{q}{2} {\cal c}_{i+1} + p {\cal c}_{i} + q {\cal c}_{i} + \frac{p}{2} {\cal c}_{i+1} + \frac{p}{2} {\cal c}_{i-1} \right].
\nonumber
\end{equation}
Various moments $\langle g_i^n \rangle$ of local mass $g_i$ can be straightforwardly calculated; see, e.g., Ref. \cite{Das}. Using the above time-evolution dynamics, the second moment of local mass $m_i(t)$ can be written as,
\begin{eqnarray}
\nonumber  \frac{d\langle g_{i}^{2}(t) \rangle}{dt} = \langle g_{i}^{2}(t)(-{\cal c}_{i} q) \rangle  + \langle g_{i+1}^{2}(t) \frac{{\cal c}_{i+1}}{2} q \rangle + \langle g_{i-1}^{2}(t) \frac{{\cal c}_{i-1}}{2} q \rangle \nonumber \\  
+ 2 \langle [g_{i}(t) g_{i+1}(t) {\cal c}_{i+1} q + g_{i}(t) g_{i-1}(t) {\cal c}_{i-1} q  - g_{i}(t) {\cal c}_{i} p] \rangle  
\nonumber  \\  
+ \frac{p}{2} \langle g_{i}(t) ({\cal c}_{i+1} + {\cal c}_{i-1})  \rangle + p \langle [{\cal c}_{i} + \frac{{\cal c}_{i+1}}{2} + \frac{{\cal c}_{i-1}}{2}] \rangle, \nonumber
\end{eqnarray}
which, in the steady state, leads to 
\begin{eqnarray}
    0 = - q \langle g_{i}^{2} \rangle + \frac{q}{2}  \langle g_{i+1}^{2} \rangle + \frac{q}{2} \langle  g_{i-1}^{2} + q \rho^2  -2 p \rho  \rangle \nonumber \\  + q \rho^2 + 2 p \rho' {\cal c}(\rho') + 2 p {\cal c}(\rho'), 
    \nonumber
\end{eqnarray}
where $\langle g_i \rangle = \rho'$ is the density in UgLHM, $\langle {\cal c}_i \rangle = {\cal c}(\rho')$ is the occupation probability of a site in UgLHM. Note that, in the above equation, we have used the following mean-field assumptions (which, as our finite-size scaling analysis indicates, could actually be exact): For $k\neq0$,
$\langle g_{i} g_{i+k} \rangle = \rho'^2 ; \langle  g_{i} {\cal c}_{i+k} \rangle = \rho' {\cal c}(\rho')$.
This particular mean-field assumption is called ``independent interval approximation'', which indeed works very well in various other mass transport processes as well. 
Upon further algebraic manipulations, we obtain occupation probability in UgLHM as a function of density $\rho'$,
 \begin{equation}
   {\cal c}(\rho') = \frac{\rho' (p - q\rho')}{p(1+\rho')}.
 \end{equation}
Then using the time-evolution of the third moment $\langle g_i^3 \rangle$ and the mean-field approximation as mentioned above, we get the expression of the second moment $\langle g_{i}^{2} \rangle$ as a function of density $\rho'$,
\begin{eqnarray}
\langle g_{i}^{2} \rangle  \equiv  \theta_{2}(\rho') = \frac{p  [1 + {\cal c}(\rho')] \rho'}{p [1 - {\cal c}(\rho')] - 2 q \rho'}.
\end{eqnarray}

\subsection{Calculation of single-site mass distribution $P(g)$ in UgLHM (equivalently, gap distribution in gLHM)}

Assuming the approximation of independent intervals (or independent gaps), we write the joint probability distribution of masses, or gaps, in a product form,
\begin{eqnarray}
{\rm Prob.}[\{g_k\}] = \frac{1}{Z_U} \prod_k w(g_k) \delta \left( \sum_k g_k - M \right),
\label{prod}
\end{eqnarray}
where masses or gaps $\{g_k\}$ are statistically independent of each other (except for the conservation constraint) and, consequently, the weight factor $w(g_k)$ for mass or gap $g_k$ at site $k$ is assumed to depend on only the gap size $g_k$ (not the neighboring gaps), $M$ is the total mass or total gap size in UgLHM and the normalization constant or the partition sum can be written as
\begin{eqnarray}
Z_U= \sum_{\{g_k\}} \prod_k w(g_k) \delta (\sum_k g_k - M). 
\label{ZU}
\end{eqnarray}
Then the probability distribution $P(g_k=g)$ of gap size $g$ can be calculated as
\begin{eqnarray}
P(g_k=g) =  \frac{w(g)}{Z_U} \sum_{\{g_k'\};k' \ne k} \prod_{k'} w(g_{k'}) \delta \left( \sum_{k'} g_{k'} - M + g \right) \nonumber
\end{eqnarray}
Provided the product form of Eq. \ref{prod} and the knowledge of the functional dependence of variance $\sigma^2(\rho') = \langle g_k^2 \rangle - \rho'^2 = \theta_{2}(\rho') - \rho'^2 = \rho'(1+\rho')(1+\rho'^2)/(1-2\rho' - \rho'^2)$ on density $\rho'$ where critical $\rho_c'=\sqrt{2}-1$ beyond which the variance is diverging, one can analytically calculate. Indeed, by following Ref. \cite{Das}, one can show that Laplace transform $\tilde w(s) = \int w(g) \exp(-s g) dg$ of weight factor $w(g)$ is related to Legendre transform of free energy density function $f_{U}(\rho')$ in UgLHM (unbounded version of the model) as given below
$$
\lambda(s) = {\rm inf}_{\rho'}\{f_U(\rho') + s \rho'\},
$$
where $\lambda(s) = - \ln \tilde w(s)$ and free energy density function $F_U(\rho')$ is calculated by integrating a fluctuation response relation $d^2 f_U/d\rho'^2=1/\sigma^2(\rho')$. The weight factor $w(g)$ can then be calculated from the inverse (discrete) Laplace transform.  As a consequence of the conductivity $\chi \sim \theta_2 = \sum g^2 P(g)$ being proportional to the second moment of gap (in gLHM as shown in the main text and also in UgLHM \cite{unbounded}) and the Einstein relation $\sigma^2=\chi/D$ (which holds also in UgLHM \cite{unbounded}), one can see that the same pole-type singular structure (a simple pole) as in the conductivity appears also in the variance $\sigma^2(\rho')$ of gap size, i.e., $\sigma^2(\rho') \sim (\rho_c-\rho')^{-1}$. This particular simple-pole singularity in  the variance $\sigma^2(\rho')$ of gap implies that the weight factor $w(g)$, for large gap sizes $g \gg 1$, must have a form of a power law \cite{Das},
\begin{eqnarray}
 w(g) \simeq C g^{-5/2},
 \label{5-by-2}
\end{eqnarray}
where $C$ is an arbitrary constant factor. Consequently, the probability distribution $P(g)$ of gap size can be written as
\begin{equation}
P(g) \equiv {\rm Prob.}(g_k=g) \sim g^{-5/2} e^{\mu(\rho')g},
\label{power-law}
\end{equation}
where $-\mu(\rho')$ is a density-dependent cut-off; here $\mu$ can be thought of as a nonequilibrium chemical potential (see the discussions in the next section). As $\rho' \rightarrow \rho_c'^{-}$ (near criticality), the chemical potential $\mu(\rho') \rightarrow 0$ and, at $\rho'=\rho_c'$ (criticality), the mass (or gap) distribution $P(g) \sim g^{-5/2}$ becomes a pure power law. Above the critical density $\rho' > \rho_c'$, the excess mass (or gap) of amount $L(\rho' - \rho_c')$ forms a condensate of gap (equivalently, a condensate of holes forms in the exclusion version of gLHM).

\subsection{Calculation of two-point correlation $c(r)$ in gLHM with $\phi(l')=\delta_{l',l}$ and $l \rightarrow \infty$}

In this section, we calculate two-point density correlation function using the mapping between gLHM and UgLHM (see main text for the mapping) and the configuration probability weight Eq. (\ref{prod}) in UgLHM.
Let us denote the weight factor $W(C)$ for a microscopic configuration $C$ in gLHM. Then the probability of a microscopic configuration $C$ can be written as 
\begin{eqnarray}
P(C) = \frac{W(C)}{Z(N,L)},
\end{eqnarray}
where $N$ and $L$ are the total number of particles and lattice sites, respectively and $Z(N,L)$ is the corresponding partition sum
\begin{eqnarray}
Z(N,L) = \sum_C W(C).
\end{eqnarray}
It is not difficult to see that the partition sums $Z_U$ in UgLHM [as in Eq. (\ref{prod})] and $Z$ in gLHM are related to each other by a simple prefactor \cite{Jain}, 
\begin{eqnarray}
Z(N,L) = \frac{L}{N} Z_U(N',L'),
\end{eqnarray}
where  $N'=L-N$ is the total mass or gap size and $L'=N$ is the number of lattice sites in UgLHM (through the mapping between gLHM and UgLHM, $N'=L-N$ and $L'=N$). The generating function (discrete Laplace transform or the ``grand-canonical'' partition sum) for the partition sum $Z_U(N',L')$ is given by
\begin{align}
\widetilde{Z}_U(z,L') = \sum_{N'=0}^{\infty} Z_U(N',L') z^{N'} = [\tilde w(z)]^{L'}
\end{align}
where $z=e^{\mu}$ is the fugacity and $\tilde w(z)$ is the generating function (discrete Laplace transform) of the weight factor $w(g)$,
$$
\tilde w(z) = \sum_{g=0}^{\infty} z^{g} w(g).
$$  
Let us now define two-point correlation function in gLHM (the exclusion version) as $c(r) = \langle n_{i}n_{i+r} \rangle - \rho^{2}$, which we calculate here by following Ref. \cite{Jain}. The first term $\langle n_{i}n_{i+r} \rangle$ in the correlation function $c(r)$ gives nonzero value when both $i$th and $(i+r)$th sites are occupied. Now consider a set of configurations in which there are $k$ holes present in between $i$th and $(i+r-1)$th site. Using the mapping between gLHM and UgLHM and summing over all such allowed configurations, we get for large $L$ and $N$ \cite{Jain},
\begin{eqnarray}
\label{cr1}
\langle n_{i} n_{i+r} \rangle & = & \sum_{k=0}^{r-1} \frac{Z_U(k,r-k) Z_U(L-N-k, N-r+k)}{Z(N,L)} \nonumber \\ &=&\rho \sum_{k=0}^{r-1} \frac{Z_U(k,r-k) Z_U(L-N-k, N-r+k)}{Z_U(L-N, N)}. \nonumber \\
\end{eqnarray}
Using Taylor series expansion,
\begin{widetext}
\begin{eqnarray}
\label{Z_Taylor}
\ln \left[ \frac{Z_U(L-N-k, N-r+k)}{Z_U(L-N, N)} \right] & = & (k-r) \frac{\partial [\ln Z_U(N'=L-N, L'=N)]}{\partial L'} - k \frac{\partial [\ln Z_U(N'=L-N, L'=N)]}{\partial N'} \nonumber \\& = & k \mu - (r-k) P 
\nonumber
\end{eqnarray}
\end{widetext}
where $\mu = \partial F(N',L')/\partial N'$ and $P = - \partial F(N',L')/\partial L'$ are chemical potential and pressure function, respectively, and $F(N',L')= -\ln Z_U(N',L')]$ is a free energy function in UgLHM, one obtains the following identity \cite{Jain},
\begin{align}
\label{cr2}
\langle n_{i} n_{i+r} \rangle = \rho e^{r\mu} \sum_{k=0}^{r} Z_U(r-k,k) e^{-k(\mu+P)}. 
\end{align}
The above identity Eq. (\ref{cr2}) can be used to obtain the generating function $G(y) = \sum_{r=0}^{\infty} y^{r} c(r)$, 
\begin{eqnarray}
G(y) &=& \left[ \frac{\rho}{1-ye^{-P} \tilde w(yz)}-\frac{\rho^{2}}{1-y} \right] \nonumber \\ & = & \left[ \frac{\rho \tilde w(z)}{\tilde w(z)-y \tilde w(yz)} - \frac{\rho^{2}}{1-y} \right],
\label{G1}
\end{eqnarray}
which is expressed here in terms of (discrete) Laplace transform $\tilde w(z)$ of the weight factor $w(g)$ and where fugacity $z = e^{\mu}$ and $e^{-P}=1/\tilde w(z)$. Note that $\tilde w(z)$ is calculated from the explicit form of the weight factor as already obtained in Eq. (\ref{5-by-2}). We now perform asymptotic analysis around the critical point $z = 1$ (i.e., $\mu =0$). By replacing the variable $y = \exp(-s)$ and then obtaining the leading order singularity in small $s=(1-y)$ expansion of Eq. (\ref{G1}), one can immediately determine the large $r$ behaviour of the correlation function $c(r)$. In the limit of small $s \rightarrow 0$, we get, from Eq. (\ref{5-by-2}), $\tilde w(yz)|_{z=1, y=1-s} \sim s^{3/2}$ and, consequently from Eq. (\ref{G1}), $G(y=1-s) \sim s^{-1/2}$. The asymptotic form of the correlation function $c(r)$ is obtained by doing inverse Laplace transform,
\begin{eqnarray}
c(r) \simeq \frac{1} {2\pi i} \int_{-i \infty}^{i \infty} ds e^{sr} G(s) \sim r^{-1/2},
\end{eqnarray}
which is precisely the functional behavior of the correlation function at criticality as mentioned in the main text. 
Note that, although there are no spatial correlations in UgLHM (unbounded version), the spatial correlations in gLHM (exclusion version) are indeed long-ranged.

\end{document}